# Perspectives of cross correlation in seismic monitoring at the International Data Centre

Dmitry Bobrov, Ivan Kitov and Lassina Zerbo

Comprehensive Nuclear-Test-Ban Treaty Organization

**Abstract**

We demonstrate that several techniques based on cross correlation are able to significantly reduce the detection threshold of seismic sources worldwide and to improve the reliability of IDC arrivals by a more accurate estimation of their defining parameters. More than ninety per cent of smaller REB events can be built in automatic processing while completely fitting the REB event definition criteria. The rate of false alarms, as compared to the events rejected from the SEL3 in the current interactive processing, has also been dramatically reduced by several powerful filters. The principal filter is the difference of arrival times between the master events and newly built events at three or more primary stations, which should lie in a narrow range of a few seconds. Two effective pre-filters are *f-k* analysis and $F_{prob}$ based on correlation traces instead of original waveforms. As a result, cross correlation may reduce the overall workload related to IDC interactive analysis and provide a precise tool for quality check for both arrivals and events.

Some major improvements in automatic and interactive processing achieved by cross correlation are illustrated by an aftershock sequence of a large continental earthquake. Exploring this sequence, we describe schematically the next steps for the development of a processing pipeline parallel to the existing IDC one in order to improve the quality of the REB together with the reduction of the magnitude threshold. The current IDC processing pipeline should be focused on the events in areas without historical seismicity which are not properly covered by REB events.

**Introduction**

The Comprehensive Nuclear-Test-Ban Treaty (CTBT) obligates each State Party not to carry out any nuclear explosions. The Technical Secretariat (TS) of the Comprehensive Nuclear-Test-Ban Treaty Organization will carry out the verification of the CTBT. The International Data Centre (IDC) is an integral part of the (currently Provisional) TS. It receives, collects, processes, analyses, reports on and archives data from the International Monitoring System (IMS). The IDC is responsible for automatic and interactive processing of the IMS data and for standard IDC products. The IDC is also required by the Treaty to progressively enhance its technical capabilities.

The methods based on cross correlation have recently shown the possibility of significant improvements in many seismological applications such as detection of low magnitude seismic events (Gibbons and Ringdal, 2006; Harris and Paik, 2006; Schaff, 2008; Schaff and Waldhauser, 2010), location of seismic events (Schaff et al., 2004; Schaff and Waldhauser, 2005; Schaff and Richards, 2011), phase identification and characterization (Gibbons, Ringdal, and Kvaerna, 2008; Harris and Dodge, 2011), event size characterization (Schaff and Richards, 2011) and event clustering (Harris and Dodge, 2011). All these improvements are of crucial importance to IDC routine processing, both automatic and interactive, and also for expert technical analysis of specific events as provided for under the Treaty. In this study, we focus on detection, phase identification, and event building in automatic processing. All events obtained by cross correlation have to be reviewed manually in accordance with the IDC rules of interactive analysis, and thus standard location algorithm was used.

The cross correlation technique can be a powerful tool for the detection of similar signals. For 3-C stations, similar signals on vertical channels may come from any azimuth and may have any slowness. The use of all three components puts some constraints on the difference between azimuth and slowness for two signals to have a high correlation coefficient. Overall, the sensitivity of a 3-C station is lower than that of a standard array station due to higher influence of ambient seismic noise. For a given level of signal, the estimates of azimuth and slowness at a 3-C station are characterized by a higher uncertainty and cross correlation may fail to distinguish between valid and wrong signals.

For array stations with many individual vertical and 3-C sensors at distances from a few hundred metres to tens of kilometres, similar signals should have similar vector slowness: the signals from different azimuths and with different apparent velocities across arrays with an aperture of several kilometres are well suppressed by destructive interference. A higher cross correlation coefficient between a given waveform and some signal from a master event is an indicator that there is a phase similar to that from the master event and this phase belongs to a source near the master event. This is a reliable method of phase identification.

Reliable detections obtained with the cross correlation technique together with efficient phase identification and event building applications, have been used to develop an independent automatic processing pipeline and to assess its performance. This technique allows a flexible approach to time windows, frequency bands, cross correlation thresholds, and other parameters controlling the overall flux and specific parameters of detections. An optimally filtered set of detections should result in a consistent and reliable list of automatically built events, which we call Cross Correlation Standard Event List, XSEL.

An effective way to test the cross correlation technique is to use a set of events which are close in time and space. A natural candidate is the aftershock sequence of a shallow event. After catastrophic earthquakes, aftershocks are distributed over a wide area and their signals are not necessarily well correlated, whereas swarms and aftershock sequences of small and moderate events usually consist of small events not recorded at teleseismic distances. It is therefore preferable to choose a sequence of an intermediate size. For an earthquake in China which occurred on 20 March 2008, we used the vertical channels of several IMS array stations to calculate cross correlation coefficients. Then, all qualified detections obtained by cross correlation were used to build all events according to the event definition criteria (EDC) adopted by the IDC for the events in the Reviewed Event Bulletin (REB). Then those events that were not found in the REB were tested by an experienced analyst according to IDC rules and guidelines.

The remainder of this paper consists of three Sections and Conclusions. Section 1 briefly presents the current scheme of automatic processing at the IDC. Selected products of the IDC are described and the Reviewed Event Bulletin is analyzed for the distribution of seismic events in space. It also describes a tentative set of processes and procedures, as related to cross correlation, and evaluates the potential improvements in the performance of automatic processing.

Section 2 introduces several basic procedures and parameters associated with the implementation of the cross correlation technique at the IDC and illustrates these procedures using two tests conducted by the Democratic People's Republic of Korea (DPRK) in 2006 and 2009. In Section 3, we present an example of a new processing pipeline as applied to the aftershock sequence of the earthquake in China. Since the cross correlation technique is constrained to distances of tens of kilometers between events and does not depend on detections from remote events, all results obtained from the analysis of this aftershock sequence can be extrapolated to the global level. The globe should be divided into a large number of intersecting cells with an optimal but small number of master events inside each cell.

In Conclusion, we touch upon some prospective methods which will be able to optimize calculations and reliability of the events built using cross correlation and estimate the level of computing capacity necessary for real-time processing.

## 1. Automatic processing at the IDC and the Reviewed Event Bulletin

According to (Coyne *et al*., 2012) the objective of seismic processing at the IDC is to produce event lists and bulletins that describe the seismic events that occurred in a given period. (We ignore here the infrasound and hydroacoustic components of the IMS which are indispensable parts of the joint seismic/infrasound/hydroacoustic automatic processing and interactive analysis.) At first, a series of Standard Event Lists (SEL) is produced by automatic processing. Then analysts review and modify the automated event list as necessary. The quality of the reviewed events is additionally checked, and all qualified events are placed in the Reviewed Event Bulletin which is the reviwed product of the IDC. It contains a list of events with their origin times, coordinates, depths, magnitudes, the relevant uncertainties and other characteristics. A large part of these characteristics is calculated in post-location (automatic) processing.

Automatic processing includes station-based processing followed by network processing. Station processing is aimed at detection of appropriate signals and estimation of their parameters, including phase identification. Network processing has to build events from the detections obtained at many stations. Event building is practically equivalent to event location when the arrival times and vector slownesses of detections (with their estimated uncertainties) at several stations fit the same source position and origin time.

At the level of station processing, the following steps are most relevant to this study: signal detection, signal azimuth and slowness estimation, and phase identification. At the IDC, detection is based on the ratio of the short-term energy to the long-term energy (STA/LTA) and the azimuth/slowness estimation uses frequency-wavenumber (*f-k*) analysis for array stations or polarization analysis for 3-C stations. We would like to extend the currently implemented procedures and techniques with those based on cross correlation. For that purpose, a number of events will be treated as master events. All seismic phases detected at IMS primary stations from these events, which should pass a rigorous quality check, will be used as waveform templates. The length of these templates depends on source-station distance and frequency band of the filters applied to original waveforms. Since the IDC is interested in the smallest possible events, waveform templates should cover only teleseismic P-waves, which are quite short (say, 2 to 4 s) for small events. When one or a few stations are available at regional distances from the event, Pn-wave and other regional phases can be included in the list of templates depending on the regional wavefield. For some regional phases, templates might be tens seconds in length.

For local and regional networks, it was demonstrated in numerous studies (e.g. Gibbons and Ringdal, 2006; Schaff and Waldhauser, 2010) that cross correlation can reduce the detection threshold by approximately one unit of magnitude, i.e. by a factor of 10 in terms of amplitude. Luckily, this gain is accompanied by a significant reduction in the uncertainty of azimuth and slowness estimates - a higher cross correlation indicates that the studied event is likely very close to the well calibrated master event for which all relevant parameters have been already accurately estimated.

Routine detection and azimuth/slowness estimation together with other tasks of station processing (data quality, improving the SNR of the data, determining amplitude and period of the signal, writing detection beams, phase association, determining the type of signal, grouping the signals at each station and identifying the phases, and location of single station events) should remain untouched. The only difference is that the set of detections obtained by these routine procedures should be reduced by those obtained by cross correlation where they have similar arrival time and other defining parameters. The reason for the exclusion of the XSEL arrivals from the IDC routine processing is the superiority of cross correlation in detection, phase identification, and azimuth/slowness estimation. However, when two signals obtained from routine and cross correlation procedures have similar arrival times but quite different azimuth, slowness, and amplitude one has to use both in the relevant processing type. Since cross correlation is very sensitive to the position of master events it will likely miss signals from the areas not yet covered by the master events.

Network processing attempts to associate a number of arrivals with the same event and then locates. Some new arrivals are sought to be associated and the arrivals already associated are tested for consistency with the new location. This process is iterated before some predefined convergence criteria are matched or stops when the number of iterations reaches its limit and the null hypothesis is rejected. The obtained locations allow converting amplitudes and periods of the arrivals in magnitudes of corresponding events, and thus estimating the event sizes. At the IDC, the Global Association (GA) applications perform all of the tasks of network processing simultaneously in several source regions and time intervals; conflicts between time intervals and regions are then resolved. For phase association, a grid-based event search procedure is applied, with the grid of cells covering the entire surface of the earth plus depth zones at which events are known historically known to occur. For each grid cell, a null hypothesis for an event is formulated and all arrivals at the nearest station are tested against the null. When an arrival at the nearest station matches the hypothesis for a given cell it is considered as a driver arrival. Using the hypothetical driver event the GA predicts the arrival times at other seismic stations. All phases within the network and the station specific predefined uncertainty bounds around the predicted arrival times, azimuth, and slowness are associated with the driver event. In automatic processing, there are several event definition criteria to be met for the event hypothesis to be confirmed. These criteria are related to the number of primary stations (one or two depending on the presence of regional phases) with arrival times and azimuth/slowness estimates within the station and cell specific uncertainty bounds. Statistically, the event definition criteria define the probability of event detection at a given rate of false alarms.

Skipping many details of network processing we would like to highlight some major features of the GA applications which can be extended or enhanced by cross correlation. The grid search covers the areas with known historical seismicity, i.e. mainly the areas with REB events. To some extent, the usage of an optimally selected subset of the REB as master events for cross correlation is equivalent to the grid search.

According to the Gutenberg-Richter law the number of events increases by a factor of 10 with magnitude falling by one unit. As a result, signals used in network processing are chiefly weak and have poor estimates of the parameters used by the GA. It should be noted that small events are the major challenge under a comprehensive Treaty – the IDC has to build all events which meet the events definition criteria for a given set of IMS data. Larger events are also of concern but they are usually easy to build and locate automatically. Cross correlation provides a more reliable detection and azimuth/slowness estimation of smaller signals and thus a more reliable association with the driver events. The cross correlation technique is very efficient in screening out all signals not matching the arrival time and vector slowness for a given master event. Therefore, the GA (or its local equivalent described in this paper) will have a much smaller, very reliable, and well prepared set of detections for network processing what will allow reducing both monitoring threshold and computational requirements. For station processing, however, computation may significantly increase. The number of master events needed for a comprehensive station and network processing can be estimated from the spatial event density in the REB and the cross correlation dependence on distance between events.

By September 2011, the REB archive contained more than 335,000 events worldwide. Some of these events are characterized by a non-zero depth, which is an effective criterion to screen out the events of natural origin; explosions are supposed to be conducted at very shallow depths. There are approximately 250,000 shallow events with depths less than 50 km. Obviously, the level of the cross correlation coefficient depends on the distance between two events. The larger is the distance the lower is the expected correlation coefficient. As shown in Section 3, for small and middle-size events, one should not expect any significant cross correlation at distances beyond 50 to 100 km. Thus, 50 km is a conservative estimate of the largest distance where cross correlation might be viable. We have calculated distances between all pairs of events from the 250,000 shallow events and constructed two important dependences on distance.

First, we have calculated the number of events which have at least one other event in various distance ranges: less than 10 km, < 20 km, < 30 km, < 40 km, and < 50 km. There is also an open ended bin for those events which have the closest neighbour beyond 50 km. We consider all events in the latter group as "isolated" events, i.e. the events without good cross correlation with any of the REB events. As mentioned above, the 50 km threshold is a tentative one and should be re-estimated according to the dependence of the correlation coefficient between all REB events on distance and the desired rate of false alarms, as obtained for the detections using cross correlation. The second distribution counts all pairs of events in the same distance ranges. The total number of such pairs may exceed the total number of events. Table 1 lists both distributions. Three figures are most important: the total number of "isolated" events, the number of events having another REB event in the range below 50 km, and the total number of pairs in the same range.

The total number of "isolated" events (3,618 from 250,000 or 1.4%) is a proxy to the expected rate of events which cannot be built by cross correlation since there is no master event within 50 km. Such events have to be built under the current framework of automatic and interactive processing, where detections are associated according to their arrival times, azimuths and slownesses obtained by standard methods. At the same time, approximately 99% of REB events would have been accurately built using cross correlation with one of the existing REB events. These are only crude estimates extrapolating the past seismicity into the future. However, with any new REB event the coverage of the globe with master events will increase and 3,618 currently isolated events may then find a pair within 50 km. The number of events having at least one event from the REB in the range below 50 km provides a very conservative estimate. Apparently, the number of events having at least one event at distance less than 50 km is 246,382.

Finally, the number of pairs spaced less than 50 km apart is a proxy to the density of events. For 246,382 events we have 30,048,847 pairs or ~122 pairs per event on average. This value allows crude estimation of the number of templates necessary to cover all REB events. Taking into account all 3,618 isolated events, ~2,100 (≈246,382/122) templates are used to cover all other events. When the correlation radius decreases, the number of necessary templates should increase.

**Table 1. Total number of events and pairs in various distance ranges**

| Range, km | <10 km | <20 km | <30 km | <40 km | <50 km | >50 km |
| --- | --- | --- | --- | --- | --- | --- |
| # events | 212,552 (85%) | 236106 (94%) | 242,412 (97%) | 245,031 (98%) | 246,382 (98.6%) | 3,618 (1.4%) |
| # pairs | 2,275,244 | 7,265,346 | 13,774,450 | 21,434,240 | 30,048,847 | - |
| # pairs/event | 10.7 | 30.8 | 56.8 | 87.5 | 122.0 | - |

Figure 1 depicts the spatial distribution of all 250,000 REB shallow events (yellow circles) and the isolated events (red circles). The overall pattern is well-known – an overwhelming majority of earthquakes is associated with plate tectonics: subduction, mid-oceanic ridges, and transform faults, i.e. seismically active zones in oceans. Within continents, even seismic events located at shallower depths are mainly associated with deeper portions of subducting plates and several active rifts.

There are two principal types of isolated event. Many events are isolated on the periphery of seismically active zones. This might be a manifestation of seismicity decaying with distance from the most active tectonic movements or the result of inaccurate location. The latter case is of importance for the cross correlation technique – the level of mislocation can be easily reduced if these events were actually close to other REB events. Therefore, one needs first to test the isolated events for cross correlation with, say, all events situated within their confidence ellipses. As a result, the number of isolated events around the seismically active zones will be dramatically reduced. This is a crucial methodological procedure to apply before one selects the optimal set of master events with their waveform templates at primary stations.

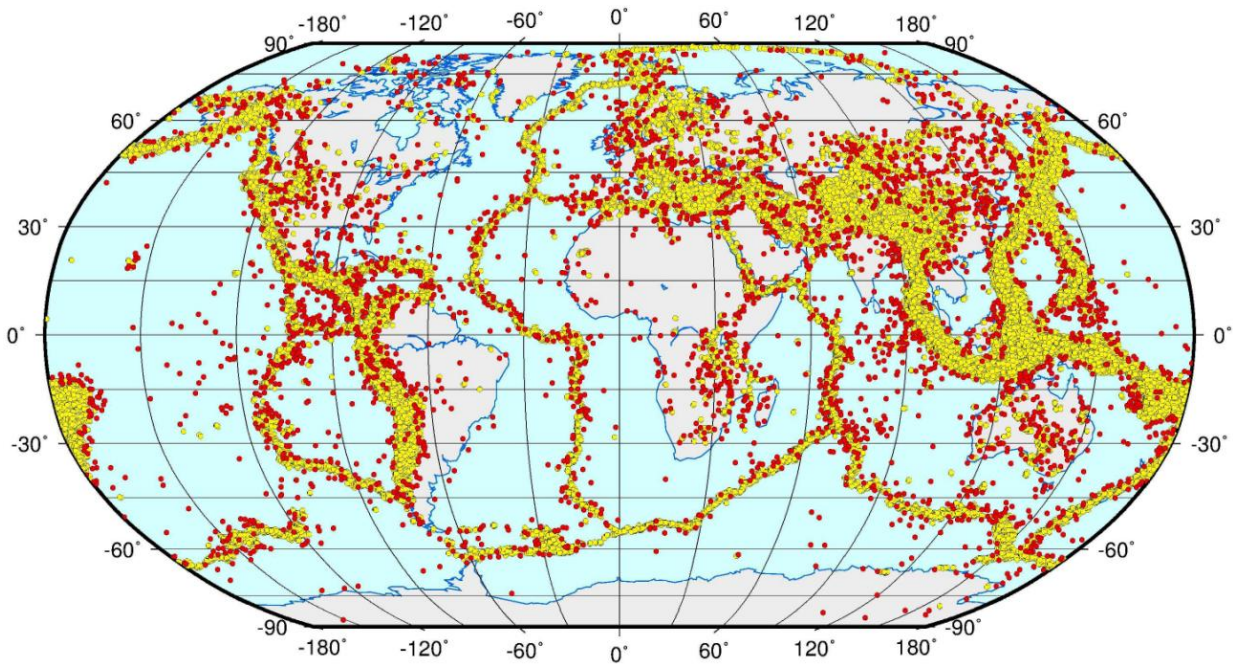

**Figure 1. Map of 3618 isolated REB events (red circles) on top of 250,000 shallow events (yellow circles).**

By definition, the set of master events has to optimally cover all seismically active areas and also include all events which have been proven to be isolated in terms of cross correlation. Hence, any new (and also past) REB event which is either in these areas or close to the isolated events has to demonstrate a predefined level of cross correlation with waveform templates for at least one master event. By putting some optimal thresholds for the correlation coefficients at a predefined number of primary stations (currently three primary stations are necessary to define a valid REB event) we can automatically build a new event. In other words, we assume that the higher correlation coefficient between waveforms from the master event and the event under study at three or more primary stations guarantees the existence of the new event and its closeness to the master event in space at a predefined level of confidence. All involved thresholds and confidence levels have to be estimated from the REB and then are subject to re-estimation in line with new information.

Waveform templates for a master event may include various types of station. The seismic part of the International Monitoring System consists of primary and auxiliary stations. The former stations are used to create initial hypotheses on seismic events which then can be corroborated and improved by data from auxiliary stations. The primary seismic stations are characterized by a continuous data flow and the auxiliary stations deliver data by request. Both primary and auxiliary stations can be either 3-C or array. Most primary stations are arrays which allow for a higher resolution of slowness and azimuth for a given P-wave arrival. (Here we consider only P-waves, which are the only type of waves recorded at teleseismic distances from small seismic events.) For cross correlation, an array station provides an additional resolution because the level of correlation between waveforms from different sources very sensitive to the time delays at the array's individual sensors. For P-waves measured at 3-C stations, only the vertical channel is useful for cross correlation. At regional distances, Sn- and Lg-waves may have larger amplitudes and all three channels are then useful for cross correlation.

In the next Section, we describe the procedures for calculation of cross correlation coefficient and several important parameters of the signals detected by cross correlation. These procedures and parameters are tested using two underground explosions conducted by the DPRK in 2006 and 2009.

## 2. Cross correlation and related techniques

Following Gibbons and Ringdal (2006), we use a normalized cross correlation function. Both time series must have the same sample rate. This condition seems to be trivial for the same seismic station but when decimation is used for reduction of the overall computation time one should be careful to use the same rate. The notation $\omega_{N,\Delta t}(t_0)$ is used to denote the discrete vector of $N$ consecutive samples of a continuous time function $\omega(t)$, where $t_0$ is the time of the first sample and $\Delta t$ is the spacing between samples:

$$\omega_{N,\Delta t}(t_0) = [\omega(t_0), \omega(t_0+\Delta t),\ldots, \omega(t_0+(N-1)\Delta t)]^T$$

The inner product of $\upsilon_{N,\Delta t}(t_\upsilon)$ and $\omega_{N,\Delta t}(t_\omega)$ is defined by

$$\langle \upsilon(t_\upsilon), \omega(t_\omega)\rangle_{N,\Delta t} = \sum_{i=0}^{N-1} \upsilon(t_\upsilon + i\Delta t)\, \omega(t_\omega + i\Delta t)$$

and the normalized cross-correlation coefficient by

$$CC[\upsilon(t_\upsilon), \omega(t_\omega)] = \frac{\langle \upsilon(t_\upsilon), \omega(t_\omega)\rangle_{N,\Delta t}}{\sqrt{\langle \upsilon(t_\upsilon), \upsilon(t_\upsilon)\rangle_{N,\Delta t}\, \langle \upsilon(t_\omega), \omega(t_\omega)\rangle_{N,\Delta t}}}$$

A good example of cross correlation as applied to seismic monitoring of underground nuclear explosions is the comparison of two tests conducted by the DPRK in 2006 (October 9) and 2009 (May 25). The 2006 event was detected by 22 IMS stations and built as REB events with $m_b$(IDC)=4.1. The 2009 event was measured by 59 IMS stations and had body wave magnitude $m_b$(IDC)=4.5. Unfortunately, the closest IMS stations USRK ($\Delta$~3.6°) and KSRS ($\Delta$~4.0°) were not operational during the first test and regional phases cannot be used for cross correlation.

Figure 2 presents two waveforms obtained at IMS station WRA from the 2009 and 2006 underground tests. These waveforms represent the origin beams, i.e. the beams with time delays between individual channels calculated with theoretical azimuth and slowness for a given source/station pair corrected for historically known biases. Both beams are filtered by a third-order band pass (Butterworth) filter between 0.8 Hz and 4.5 Hz. The peak-to-peak amplitude ratio measured from these seismograms is approximately 4. The filtered waveforms are similar and thus have a high cross correlation coefficient. We have selected a 6 s window for cross correlation with a 1 s lead and 5 s signal segment. The estimated value of cross correlation coefficient is 0.85.

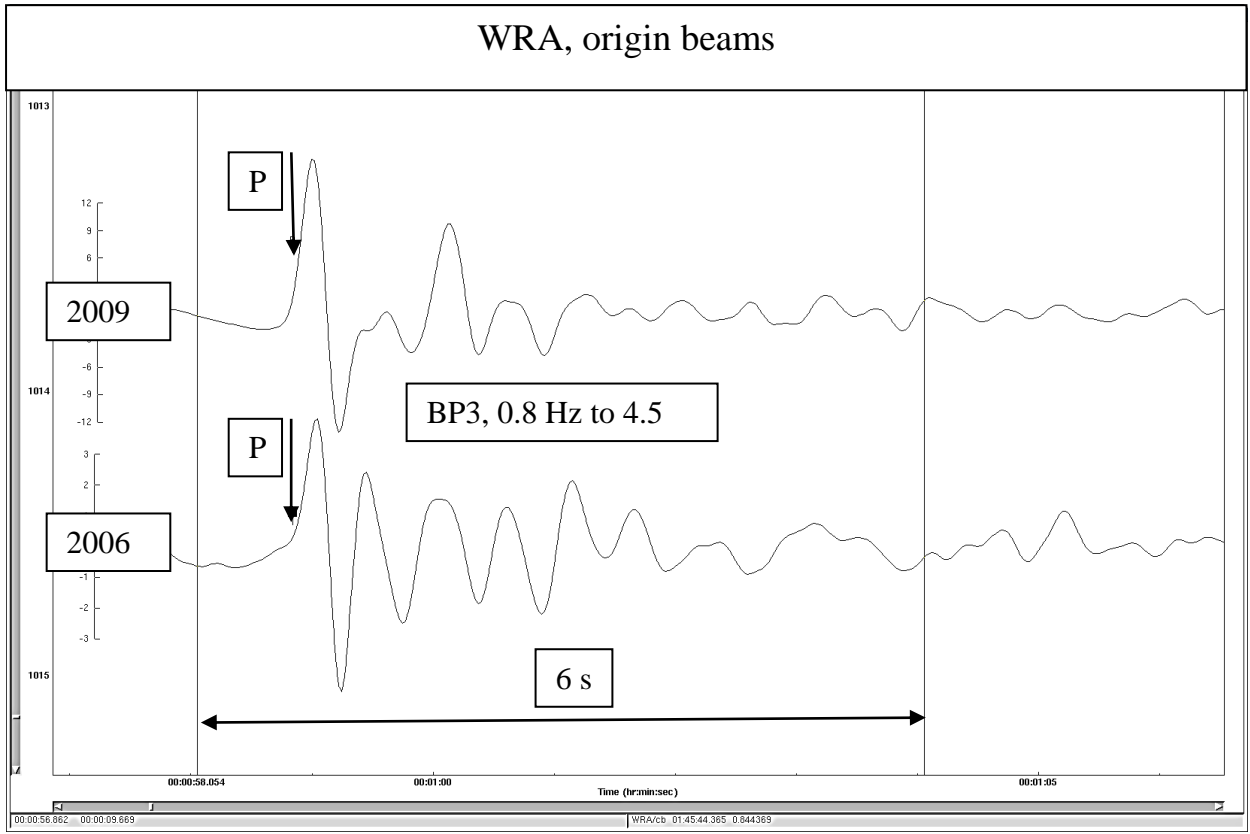

**Figure 2. Comparison of two origin beams at station WRA from the 2006 and 2009 tests conducted in the DPRK. Both beams are filtered by a third-order band pass filter between 0.8 Hz and 4.5 Hz. The filtered waveforms are similar. We have selected a 6 s window for cross correlation with a 1 s lead and 5 s signal segments.**

The origin beams, however, do not provide the best waveforms for the estimation of cross correlation coefficients. There are tangible beam losses associated with the difference between theoretical (used in the origin beams) and actual arrival times at individual sensors of a given array station. Actual signals in the origin beam are not properly synchronized and suffer some destructive interference, which is appropriate for suppression of microseismic noise but not signals. Therefore, the use of all individual waveforms, as they are, in the estimation of cross correlation coefficient is superior to any beam – they include true time delays between channels.

Figure 3 illustrates the procedure of cross correlation between a template and waveform. Here, we use continuous waveforms recorded by twenty four vertical channels of IMS station WRA and a six second long template representing a clear signal. One cross correlation coefficient is calculated over the entire template with all channels aligned (in the order of station names) in one time series of 24x6 s. This coefficient is associated with the absolute time of the first point in the waveform (usually referred to the reference station of the array under investigation). In the template, individual channels are shifted in time according to theoretical travel time residuals defined by azimuth and slowness of the origin (source/receiver) beam. (As clear from the Figure, these theoretical time delays do not guarantee signal synchronization between channels.) In the cross correlated waveform, all individual channels are also shifted by the same time delays as in the template. Hence, the empirical (true) time delays between the channels are retained when the template is convolved with the waveforms.

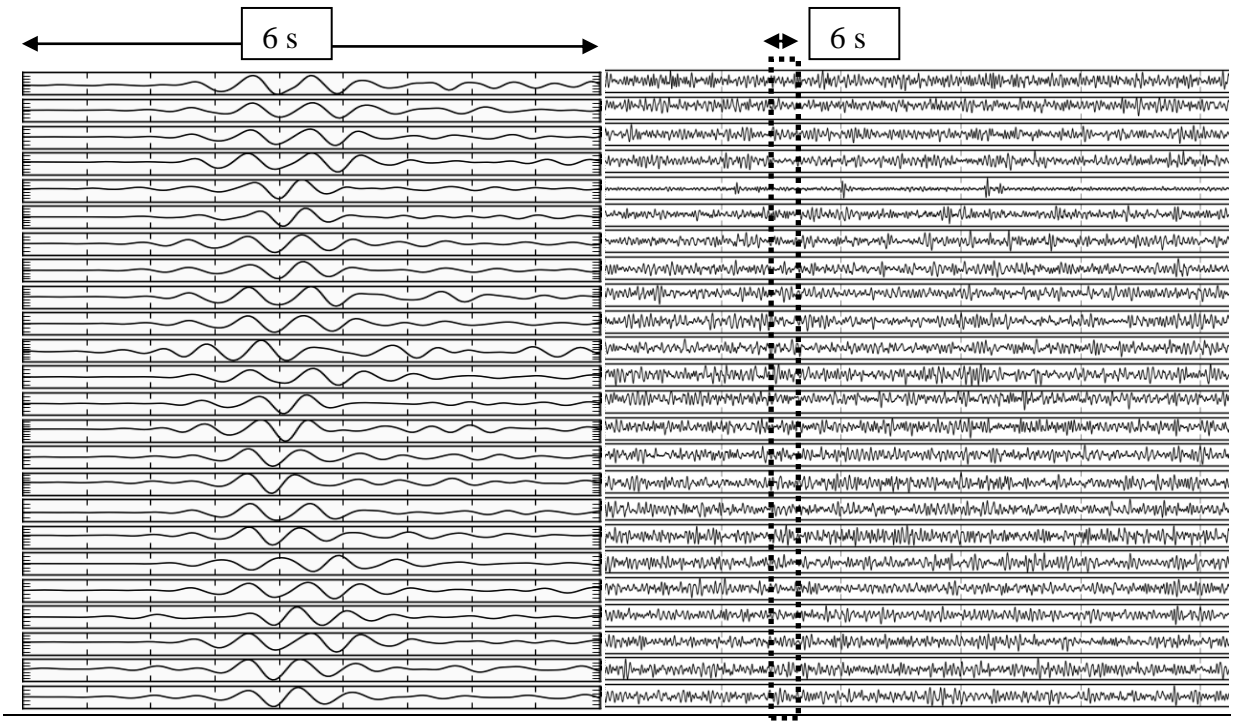

**Figure 3. IMS seismic array WRA, a template and continuous waveforms. One correlation coefficient is calculated over the entire template with all channels aligned in one record. In the template, individual channels are shifted in time according to theoretical travel time residuals defined by azimuth and slowness of the origin (source/receiver) beam. All continuous waveforms are also shifted by the same time delays between individual channels.**

The length of seismic signal recorded at teleseismic distances depends on magnitude: smaller events are characterized by shorter visible signals. In addition, smaller earthquakes produce signals enriched by higher frequencies due to higher corner frequencies of their sources. Here, we are looking for the smallest events which are likely to be missed by standard detection algorithms and event building tools used at the IDC. Therefore, the length of template widows should not be large, and should only include valid signals from small and moderate-size events.

The difference in frequency content of microseismic noise and signals at IMS stations requires a number of filters covering the whole spectral range of seismic signals from 0.8 Hz to 6 Hz.

Table 2 lists time windows and frequency bands of eight templates used in this study. All templates include several seconds of P- or Pn-wave signal and a short time interval before the signal (lead), which provides additional flexibility in onset time. (At the initial stage of our study, we do not include the PKP phase, which is a primary phase at distances beyond ~115$^o$ and might be of importance for specific seismic regions.) The length of a given window depends on its frequency band. For the low-frequency (BP, order 3) filter between 0.8 Hz and 2.0 Hz, the length is 6.5 s and includes 1 s before the arrival time. For the high-frequency filter between 3 Hz and 6 Hz, the length is only 4.5 s. For Pn-waves, the length is 11 s and does not depend on frequency. The Pn templates also include 1 s of preceding noise.

**Table 2. Time windows and frequency bands of the templates for P- and Pn-waves**

| Phase | Filter | | | | Window, s | | |
|---|---|---|---|---|---|---|---|
| | Low (Hz) | High (Hz) | Type | order | Lead | Signal | Name |
| P | 0.8 | 2.0 | BP | 3 | 1.0 | 5.5 | P0820 |
| P | 1.5 | 3.0 | BP | 3 | 1.0 | 4.5 | P1530 |
| P | 2.0 | 4.0 | BP | 3 | 1.0 | 3.5 | P2040 |
| P | 3.0 | 6.0 | BP | 3 | 1.0 | 3.5 | P3060 |
| Pn | 0.8 | 2.0 | BP | 3 | 1.0 | 10.0 | Pn0820 |
| Pn | 1.5 | 3.0 | BP | 3 | 1.0 | 10.0 | Pn1530 |
| Pn | 2.0 | 4.0 | BP | 3 | 1.0 | 10.0 | Pn2040 |
| Pn | 3.0 | 6.0 | BP | 3 | 1.0 | 10.0 | Pn3060 |

When a time series of cross correlation coefficients is calculated for a given interval, say two hours, one can apply signal detection algorithms. There is no theoretically justified unique $CC$ threshold, $CC_{tr}$, used to define a new arrival; rather appropriate thresholds should be determined empirically. These thresholds are likely to be station dependent and vary with geographical coordinates and source depth. For shallow events, free surface reflections may introduce varying interference patterns.

For two neighbouring events, the level of cross correlation coefficient depends on the distance between them and the similarity of source functions, as well as upon signal frequency. As a rule, the larger the distance, the lower the corresponding $CC$ as caused by degrading coherency of signals on various channels. The similarity of source functions can also deteriorate with the difference in magnitude, especially for short time windows used in our templates. Shallow earthquakes usually generate emergent signals. For larger events, an early (and different in shape) part of the signal from the same location may emerge from the noise and thus be used in corresponding templates, while it is not seen and used from smaller events. As a result, the level of cross correlation may decrease even for collocated events.

For weak signals, the absolute level of correlation coefficient for collocated events can be reduced by the effect of uncorrelated seismic noise mixed with the signals. Hence, before using *CC* as a detector, one has to enhance the detection procedure. There are many possibilities and likely the simplest one is the STA/LTA detector, which is already implemented at the IDC for original waveforms. This detector is based on a running short-term-average (STA) and long-term-average (LTA), which is computed recursively using previously computed STA values. The LTA lags behind the STA by a half of the STA window. For a time series *x(n)*, where *n* is the sample index and *x(n)* is the amplitude at sample *n*, the initial value of the STA, *stav*, is calculated as

$$stav\left(\frac{S}{2}\right) = \frac{1}{S}\sum_{s=0}^{S-1}|x(s)|$$

where *S* is the number of samples in the STA window. Recursion is used to compute consequent values of the STA:

$$stav(k) = stav(k-1) + \frac{1}{S}\left[\left|x\left(k+\frac{S}{2}\right)\right| - \left|x\left(k-1-\frac{S}{2}\right)\right|\right]$$

where $(S/2) \leq k \leq (N-1) - (S/2)$, and *N* is the number of available samples in the time series. For the end-segment intervals, $k \leq S/2$ and $\geq (N-1) - (S/2)$, $stav(k) = stav(S/2)$ and $stav(k) = stav((N-1) - S/2)$, respectively.

The LTA, *ltav(k)*, is computed recursively from the previous STA:

$$ltav(k) = \left(1 - \frac{1}{L}\right)ltav(k-1) + \frac{1}{L}stav(k-S)$$

where *L* is the number of samples in the LTA window. The length of the STA and LTA windows have to be defined empirically as associated with spectral properties of seismic noise and expected signal. We have carried out a brief investigation and determined the following windows: 0.8 s for the STA and 20 s for the LTA.

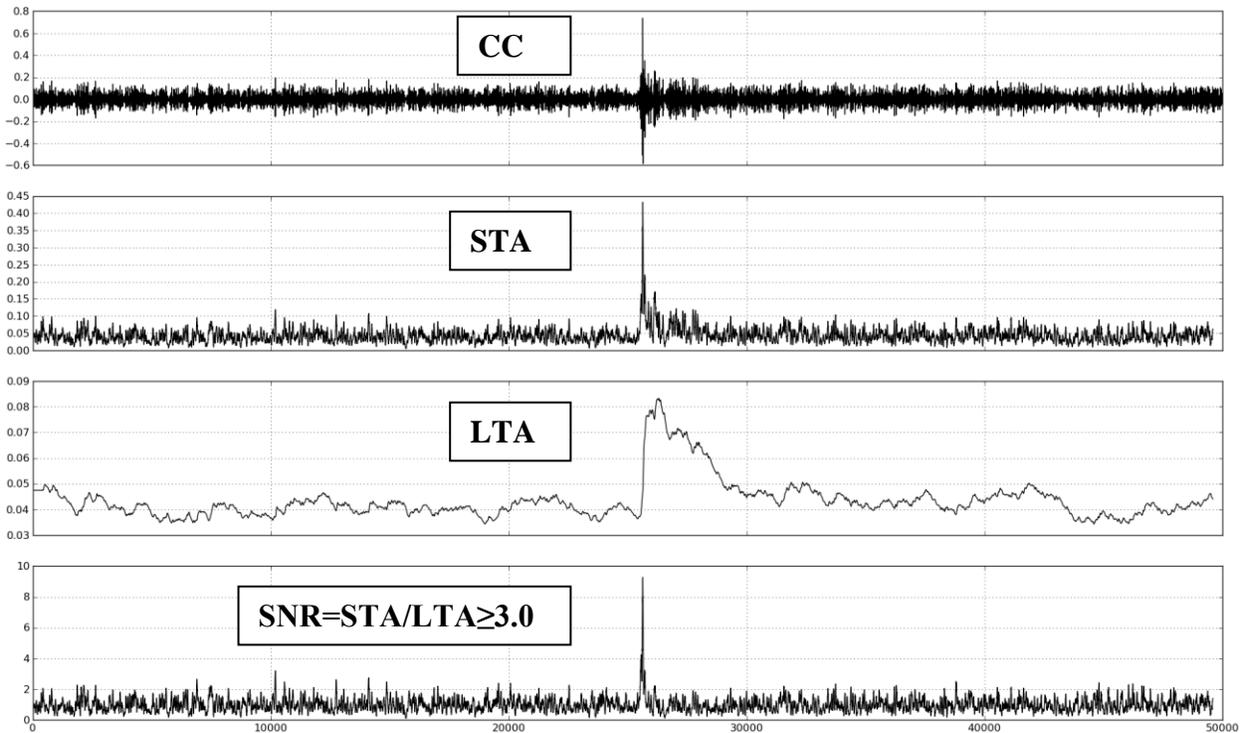

**Figure 4. An example of cross correlation analysis as applied to the 2009 and 2006 DPRK tests.**

Figure 4 illustrates the STA/LTA detection process. In the upper panel, a *CC* time series is shown as obtained by cross correlation of the template of the 2009 DPRK explosion and a 42-minute time window centered at the arrival of the 2006 DPRK test at station AKASG. The next two panels demonstrate the relevant STA and LTA. In the lower panel, the STA/LTA ratio is shown which defines the signal-to-noise ratio, $SNR_{CC}$. A valid signal is detected when the level of $SNR_{CC}$ is above 3.0. This is a tentative but conservative threshold. Before we gather a statistically significant set of arrivals and test them manually it would be premature to reduce the detection threshold from cross correlation traces. It is worth noting that for original waveforms a valid signal usually has SNR>2.0, but the *CC* detector can find a valid signal with standard

SNR=0.7. The latter value is put as a preliminary SNR threshold as calculated from the original waveforms. All in all, there is only one valid signal in Figure 4, which undoubtedly belongs to the 2006 event.

Seismic measurements are not always perfect and a researcher may face numerous data problems such as spikes, gaps, high noise at a few channels, and wrong polarity in actual waveforms. Among these problems, only wrong polarity is not a challenge for the estimates of cross correlation coefficient. Spikes and short gaps (up to 5 points) in data can be effectively suppressed and recovered by various interpolation methods used at the IDC. Longer data gaps are usually masked and the relevant readings are not used in automatic IDC processing. For cross correlation, these segments of masked data may introduce artificial steps in the *CC* time series due to a sudden change in the number of used channels. Such steps might be wrongly interpreted by the STA/LTA detector as signals. In order to reduce the influence of these masked data gaps we have introduced a cross correlation coefficient averaged over all working channels, which is called beam CC, *BCC*.

There are several IMS primary stations which detected both DPRK events. We have calculated cross correlation coefficients for both events as master ones. There is a slight difference in time delays between individual channels in the relevant waveform templates. These delays are responsible for the difference between cross correlation coefficients when the master and slave events are swopped. Table 3 lists all peak *CC*s for both nuclear tests and proves that the second event could be easily built automatically using cross correlation in line with the distance of several kilometers between the relevant IDC locations. Moreover, the cross correlation detector has demonstrated its efficiency by finding an additional P-wave arrival at station ARCES from the 2006 events which had been missed in automatic and interactive processing. The 2009 template for ARCES has been used.

Generally, a higher cross correlation between two waveforms at one station is a very reliable indication of the spatial closeness between their sources. However, there are a few cases when cross correlation is high for distant events, with the master event being much smaller than that obtained by cross correlation. There are several methods to remove such spurious correlations. One can consider these methods as additional filters applied to the flux of detections used for event building. IDC automatic processing uses *f-k* analysis, as applied to the original waveforms, in order to estimate the difference in slowness and azimuth between two events.

When this difference is high one can reject the null hypothesis that these events are close. The overall resolution of *f-k* analysis is not very high, especially, for small events generating weak signals which are a major concern for the IDC.

Gibbons *et al.* (2008) proposed to apply *f-k* analysis to the cross correlation time series. This allows a significant improvement in the overall resolution due to the sensitivity of correlation to the distance between events and effective noise suppression. Thus, we calculate azimuth and slowness using *f-k* for correlation time series for all detections obtained by cross correlation. This procedure allows effectively rejecting the cross correlation detections from remote events.

In *f-k* analysis, spectra are computed from the traces of cross correlation coefficients calculated for individual channels. For each slowness vector, the *f-k* power spectrum, $P(S_n, S_e)$, is calculated as:

$$P(S_n, S_e) = \frac{\sum_{f=f_1}^{f_2} \left| \sum_{i=1}^{J} F_i(f) \cdot e^{2\pi\sqrt{-1} f (S_n \cdot dnorth_i + S_e \cdot deast_i)} \right|^2}{J \cdot \sum_{f=f_1}^{f_2} \left\{ \sum_{i=1}^{J} F_i(f)^2 \right\}}$$

where $F_i(f)$ is the Fourier amplitude of the *i*th trace of cross correlation coefficient at frequency *f*, $S_e$ and $S_n$ are the east-west and north-south components of the vector slowness, $deast_i$ and $dnorth_i$ are the east-west and north-south coordinates, respectively, of the *i*th sensor array element relative to the reference station, $f_1$ and $f_2$ are the low and high frequency limits, *J* is the number of sensor elements in the array. The sum $S_e \cdot deast_i + S_n \cdot dnorth_i$ defines the theoretical time delay between the reference station and the *i*th sensor element. These delays are used in the templates. The slowness coordinates of the peaks in the *f-k* power spectrum are used to calculate the azimuth and slowness of the signal's spatially coherent plane wave energy. When the deviation from the master's azimuth and slowness is above some predefined thresholds the signal under investigation is not considered for event building.

Another method of pre-selection of appropriate arrivals can be based on the ratio of norms used for cross correlation of a master and slave events. Gibbons and Ringdal (2006) introduced an amplitude scaling factor:

$\alpha = \boldsymbol{x} \cdot \boldsymbol{y} / \boldsymbol{x} \cdot \boldsymbol{x}$

where *x* and *y* are the vectors of data for the master and slave event, respectively. For two collocated events with the same source time history but different amplitudes, the amplitude scaling factor completely defines the difference in sizes. For close events with similar source functions, the amplitude scaling factor defines the least square solution of the equation *y = ax+n*, where *n* is uncorrelated noise.

Schaff and Richards (2011) noticed that $\alpha$ is equivalent to the unnormalized cross correlation coefficient divided by the inner product of the master waveform. For close events, they defined a relative magnitude as the logarithm of α and demonstrated a significant (~5 times) reduction in the variance in this magnitude relative to the conventional magnitude obtained from the catalogue. Hence, for detections obtained by cross correlation, the relative scaling factor can be a more reliable screening criterion in the automatic event building than the currently used magnitude criterion. It should be noticed that the difference in station magnitudes is a very effective (dynamic) parameter which resolves a substantial proportion of the conflicts between events in the GA when they share the same phase by time, azimuth, and slowness.

For a given event, the relative scaling factors, as based on the same master event at different stations have to be close. For the current configuration adopted in automatic processing, for a phase to be associated with a given event, the difference between station- and network-averaged magnitudes cannot exceed two units of magnitude. For the relative scaling factor, this difference should not exceed 10 to 20 times. Since the logarithm of α is equivalent to body wave magnitude, the difference in log$\alpha$ should not exceed 1.3.

The relative scaling factor is not the best measure of relative event sizes, however. It includes the cross correlation coefficient which depends on the distance between events. For very close events and *CC*~1.0, there is almost no bias introduced by *CC*. For the events with *CC* varying between 0.2 and 1.0 because of distance the bias in logα may be substantial and the overall dispersion at various stations may grow significantly. In order to reduce the influence of the distance, we propose to use the ratio of norms │*x*│/│*y*│ instead of α. The logarithm of the ratio, *RM* = log(│*x*│/│*y*│) = log│*x*│ - log│*y*│, is essentially the magnitude difference (or relative magnitude, *RM*) between two events, where the magnitude is based on the RMS amplitude in the template time window instead of the peak-to-peak amplitude. This difference has a clear physical meaning for close events with similar waveforms, i.e. for events with a higher level of cross correlation. It does not work for remote events because their propagation paths and source functions are quite different and one has to introduce a standard magnitude scale. Since all variations related to *CC* are excluded, *RM* fluctuates less across the stations measured both events than log$\alpha$, as shown in Table 3 for two explosions conducted by the DPRK. Hence, *RM* is the best dynamic parameter for discrimination between genuine and dynamically inappropriate arrivals for a given event at several stations. The decision line for *RM* has to be determined empirically from the whole set of REB events matching high quality criteria, say, *CC*>0.2 and SNR$_{CC}$> 3.0 for all array stations.

**Table 3. Cross correlation coefficients at IMS primary stations for the 2006 and 2009 tests used as master events**

| STA | Master | Phase | Filter | CC | $\alpha$ | RM |
|---|---|---|---|---|---|---|
| AKASG | 2009 | P | P1530 | 0.666 | -0.740 | 0.564 |
| AKASG | 2006 | P | P1530 | 0.674 | 0.393 | -0.564 |
| ASAR | 2009 | P | P1530 | 0.660 | -0.652 | 0.471 |
| ASAR | 2006 | P | P1530 | 0.674 | 0.300 | -0.472 |
| GERES | 2009 | P | P1530 | 0.571 | -0.818 | 0.575 |
| GERES | 2006 | P | P1530 | 0.549 | 0.311 | -0.571 |
| MJAR | 2009 | Pn | Pn2040 | 0.677 | -0.479 | 0.309 |
| MJAR | 2006 | Pn | Pn2040 | 0.685 | 0.145 | -0.309 |
| MKAR | 2009 | P | P0820 | 0.522 | -0.952 | 0.670 |
| MKAR | 2006 | P | P0820 | 0.517 | 0.374 | -0.661 |
| NOA | 2009 | P | P2040 | -0.757 | -0.602 | 0.481 |
| NOA | 2006 | P | P2040 | -0.758 | 0.365 | -0.485 |
| NVAR | 2009 | P | P1530 | 0.957 | -0.590 | 0.571 |
| NVAR | 2006 | P | P1530 | 0.956 | 0.553 | -0.573 |
| PDAR | 2009 | P | P0820 | 0.759 | -0.697 | 0.578 |
| PDAR | 2006 | P | P0820 | 0.753 | 0.455 | -0.578 |
| SONM | 2009 | Pn | Pn0820 | 0.617 | -0.719 | 0.509 |
| SONM | 2006 | Pn | Pn0820 | 0.630 | 0.317 | -0.518 |
| WRA | 2009 | P | P1530 | -0.903 | -0.462 | 0.417 |
| WRA | 2006 | P | P1530 | -0.907 | 0.374 | -0.416 |

The relative magnitude can be extrapolated to the global level since the seismicity is practically continuous in terms of cross correlation. From Figure 1 one can judge that almost any two events, with the exception of those isolated within continents, can be connected through a chain of a few master events. Therefore, one will be able to balance relative magnitudes *RM* over all chains of neighbouring master events. The extrapolation of the relative magnitude at the global level is of an extraordinary importance for seismic monitoring. When connected by quantitative relationships with the global scale of body wave magnitude, the globalized relative magnitude will allow increasing the accuracy and reducing the uncertainty of the body wave magnitude estimates worldwide. It may significantly improve the performance of the screening criterion based on the difference between $m_b$ and $M_s$.

Table 3 includes the estimates of $\log\alpha$ and *RM* for all stations where both events were detected. The standard deviation of $\log\alpha$ for ten stations is 0.15 and only 0.1 for *RM*. Hence, the relative magnitude based on the norms of signals in (the same) template windows has a smaller variance and is preferable for discrimination in event building.

**Table 4. Travel time, azimuth and slowness residuals at IMS primary stations for the 2006 and 2009 tests used as master events.**

| STA | Master | $t_{res}$ | $CC\_az_{res}$ | $az_{res}$ | $CC\_slo_{res}$ | $slo_{res}$ |
|---|---|---|---|---|---|---|
| AKASG | 2009 | -0.077 | -0.07 | -3.00 | -0.20 | 0.25 |
| AKASG | 2006 | 0.075 | -0.54 | -3.10 | 0.25 | 0.10 |
| ASAR | 2009 | -0.154 | 2.03 | 2.00 | 0.15 | 1.40 |
| ASAR | 2006 | 0.178 | -2.65 | 5.70 | -0.04 | 0.29 |
| GERES | 2009 | 0.210 | 4.46 | -12.50 | -0.15 | 0.12 |
| GERES | 2006 | -0.212 | -0.42 | -10.60 | 1.33 | 0.12 |
| MJAR | 2009 | 0.147 | -0.92 | -5.90 | -0.12 | 1.31 |
| MJAR | 2006 | -0.137 | 1.93 | -4.00 | -0.06 | -0.47 |
| MKAR | 2009 | 0.716 | -0.46 | 8.60 | -2.45 | 2.23 |
| MKAR | 2006 | -0.718 | 3.15 | 6.50 | -0.04 | 0.82 |
| NOA | 2009 | -0.402 | 8.66 | -0.10 | 1.55 | 0.24 |
| NOA | 2006 | 0.400 | 4.08 | -0.80 | -0.25 | 0.27 |
| NVAR | 2009 | 0.098 | -0.26 | -2.00 | 0.47 | 1.08 |
| NVAR | 2006 | -0.100 | -3.33 | -5.30 | -0.05 | 0.49 |
| PDAR | 2009 | 0.048 | 2.88 | 18.20 | -1.52 | -1.80 |
| PDAR | 2006 | -0.025 | -29.79 | -2.40 | 1.25 | -1.25 |
| SONM | 2009 | 0.558 | 3.53 | 6.20 | -0.58 | 0.24 |
| SONM | 2006 | -0.556 | -1.21 | 4.80 | 1.39 | -0.24 |
| WRA | 2009 | 0.242 | -0.28 | 1.30 | 0.05 | 0.46 |
| WRA | 2006 | -0.24451 | -1.03 | 2.10 | 0.05 | 0.37 |

Table 4 demonstrates the extremely accurate estimates of arrival time residuals, $t_{res}$, azimuth residuals, $az_{res}$, and slowness residuals, $slo_{res}$. When obtained by standard *f-k* analysis applied to cross correlation time series instead of original waveforms, these residuals (marked as CC_*) are, chiefly, smaller by a factor of 2 and more. It is worth noting that all these estimates were made in automatic processing.

As a complementary study, we have carried out a thorough search for smaller aftershocks of the 2009 event. We have processed seismic data from primary stations five days after the event. Both nuclear tests were used as master events; the Pn-waves at stations USRK and KSRS from the 2009 event have been also used as waveform templates. There was no REB-ready event near the epicenter of the second test. Taking into account the level of correlation between the nuclear tests, we would estimate body wave magnitude of the largest possible aftershock as $m_b$(IDC)~2.5. For a bigger aftershock collocated with the 2009 event, the cross correlation coefficient has to be above the threshold adopted in our study (*CC*>0.2) at a few primary stations.

Tables 3 and 4 both evidence that the cross correlation and *f-k* analysis provide a reliable tool for detection of signals from close events and accurate estimates of their arrival time, azimuth and slowness. Due to the shape similarity, one can use the ratio of the RMS amplitudes in a predefined time window (and the same frequency band) as a robust and station independent characteristic of the relative event sizes. Overall, these findings provide a solid basis for automatic detection and event building.

In Section 3, we describe principal details of a new processing pipeline as based on cross correlation and present some preliminary results. We have carried out a feasibility study and obtained a tentative XSEL for an aftershock sequence of a large continental earthquake where we presumed no historical seismicity. Therefore, standard automatic and interactive processing is needed to populate the set of master events before any cross correlation techniques can be applied. This might be a common feature for a few intra-continental earthquakes and likely for underground nuclear explosions. Cross correlation might be less efficient in the areas without historical seismicity and thus is not able to completely replace standard IDC processing.

### 3. Testing cross correlation procedures

In Section 2, we have described general features of cross correlation as a detector and also introduced a few non-standard parameters for the cross correlation detections together with the relevant estimation procedures. All standard parameters for seismic phases adopted by the IDC are also estimated for all arrivals. Two underground tests conducted by the DPRK have demonstrated the efficiency of the detector and the reliability of such parameters as arrival time, azimuth and slowness as estimated from the traces of cross correlation coefficient as well as relative magnitude.

The next natural step is to recover a sequence of seismic events from a small area (say, from 50 km to 100 km in radius) using master events. Having a flux of detections obtained by cross correlation from a given master event or a set of master events one has to process them in an automatic pipeline to build a cross correlation standard event list (XSEL). Thus, in this Section we delineate the pipeline and also tune the parameters controlling the flux of detections and the quality of XSEL events. This is a feasibility study.

We have chosen a large earthquake in China that occurred at 22:32:56 on 20 March 2008. This earthquake was detected by many primary and auxiliary IMS stations and starting from the automatic location an event was built by IDC analysts with body wave magnitude $m_b$(IDC)= 5.41. The event had a short but prominent aftershock sequence also recorded by the IMS. There are 142 events (the main shock counted in) in the Reviewed Event Bulletin during five days after the earthquake and within 100 km from the main shock. Cross correlation is a technique which requires extensive computation resources and we have limited our testing to five full days including the whole day of the main shock.

The aftershocks located by IDC provide several opportunities to test cross correlation as a method for signal detection, phase identification, and event building. First, we must endeavor to reproduce the existing REB using cross correlation. Secondly, we have to check the REB aftershocks for internal consistency in terms of cross correlation and populate the list of master events by selecting those events which provide the largest number of good cross correlations. The aftershocks in the REB without any master event, i.e. those events which do not show appropriate cross correlation with any other aftershock in the sequence, have to be checked manually for consistency and/or for wrong phase associations. In a sense, this check for consistency is a part of quality check which should be incorporated in interactive processing. Thirdly, we should search for new events (not in the REB) in the same time slot which match the IDC event definition criteria. In short, the EDC require that an REB event has to be detected by at least three IMS primary stations with arrival time, azimuth and slowness within the station and phase specific uncertainty bounds (Coyne *et al.*, 2012). Fourthly, we must determine those parameters of detection and event building procedures which provide the highest resolution with a predefined false alarm rate. For interactive processing, this rate should be relatively low. The success of these parameters estimation depends on many factors including time, spatial and magnitude distribution of events. The parameters obtained for the studied aftershock sequence cannot be transported to other areas and from station to station as they are. Any individual region/station pair deserves an independent estimation of all controlling parameters.

The studied aftershock sequence was built by IDC analysts. For the purposes of our study, we have all REB events in order to select the best set of master events in order to reproduce the entire sequence. As a start point, we use all 142 REB events within 100 km of the main shock. These events have magnitudes $m_b$(IDC) between 2.84 and 5.41 and include from 3 to 14 primary array stations. This exercise should provide a complete set of information on cross correlation between the events created in an interactive mode in a previously unknown area. Then the best subset has to be retrieved which should be used to process data in an automatic mode to find all valid REB and new events within the correlation radius. It is presumed that all new events obtained in automatic processing have to be reviewed by analysts, before migrated into the XSEL. These new events may also be tested as potential master events.

There is a trade-off between the size of a master event and its efficiency for cross correlation. For many earthquakes, it takes several seconds for a signal to reach its peak value as related to source mechanism and propagation path. For smaller events, the initial portion of their signals at IMS stations may not exceed the level of microseismic noise, and thus it is likely not included in waveform templates. For larger events (> 5.0), the whole signal is usually above the noise level, and thus includes the initial few seconds missed in the small event's templates. The waveform template for a larger event might not find a similar signal from a collocated but much smaller event (which is below the noise level) and cross correlation fails. At the same time, the waveform template for the smaller event does contain signals repeating some later segments from the large event.

On the other hand, larger events contain more stations and their signals have much higher SNR. Generally, they provide better templates for cross correlation which are not spoiled by seismic noise. If the difference in shape between the template of a big event and the signal from a collocated but much smaller event is not large then the time delays between individual sensors define the level of cross correlation. In this case, the master event with large magnitudes can be effective in finding of smaller sources. In order to mitigate the risk of missing smallest events one can use collocated events with different magnitudes as master events. The only requirement for a small event to be a master one is clear waveform templates at many stations.

The primary IMS network includes many array stations and few 3-C stations. Theoretically, an array allows signal amplification proportional to the square root of the number of elements, when signals are spatially well-correlated and noise is incoherent. Therefore, array stations are more efficient in detection of weaker signals and smaller events world-wide. However, all 3-C stations are very important for detection of low-amplitude signals when no array stations are available at local and regional distances. Without loss of generality, we use only primary array stations in this study. There is no primary 3-C station at regional distances from the aftershock sequence. Auxiliary IMS stations provide data only by request; their waveforms are not continuous and thus not used.

Figure 5 depicts a map of IMS stations with their roles in the study. For the studied aftershock sequence, thirteen primary array stations at teleseismic distances reported P-wave detections: AKASG, ARCES, CMAR, FINES, GERES, KSRS, MJAR, NOA, PETK, SONM, WRA, YKA, and ZALV. There is one primary station at a distance below $18^o$ - MKAR. It regularly reported detections of the Pn-phase. There are also two auxiliary arrays – BVAR and KURK which could be used for cross correlation estimates. When continuous waveforms from these stations are available they can be included in templates.

Overall, we have designed the following tentative procedure for event building. As a first step, cross correlation coefficients have to be calculated for all 142 earthquakes as master events through the entire five-day-long records. For a given master event, $CC$ are calculated only for time defining P-waves (Pn-wave for MKAR) with SNR>2.0. One aftershock with $m_b$(IDC)=3.11 has three primary stations but only two P-waves with SNR>2.0. Low SNR is equivalent to poor signals and thus poor templates. Since an event with two primary stations (two templates) cannot find a valid REB event (at least 3 primary stations are needed) we did not process this aftershock as a master even. As a result, the total number of master events is 141.

For a given multichannel template, the cross correlation coefficient is a time series of the same length at the original waveform. When the $CC$ and its $SNR_{CC}$ exceed some predefined thresholds (these station/region dependent thresholds have to be determined empirically) an arrival is written into a database and several parameters are calculated from the cross correlation traces (e.g. arrival time, azimuth, and slowness) and original waveforms (e.g. amplitude, period, and *RM*).

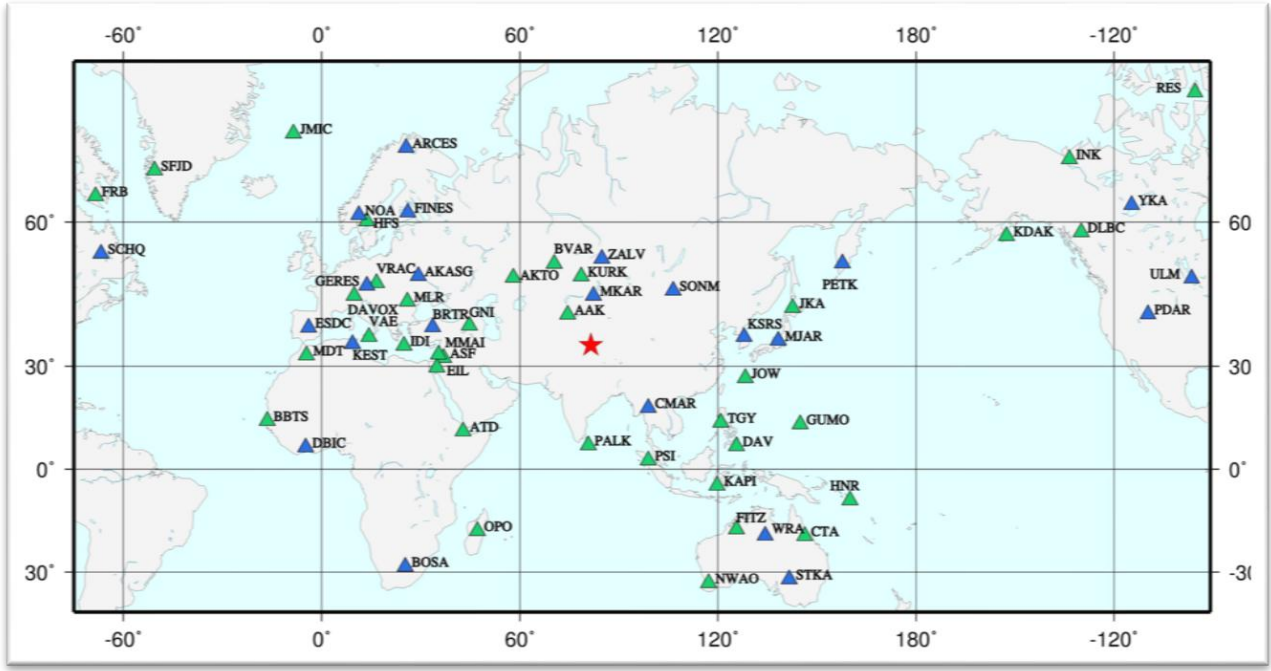

**Figure 5. Primary (blue) and auxiliary (green) IMS stations reported Pn- and P-waves from the main shock (red star).**

For a station in a master event, we use four templates with different frequency bands and time windows (see Table 2) to generate a single time series of cross correlation coefficient. It may so happen that there are two or more *CC* detections from different templates within several seconds and we have to select only one of them. Then we seek for the largest *CC* value among all detections within a four-second interval and attribute its value to the sought detection in the *CC* coefficient series. The onset time and all other parameters (e.g. azimuth, slowness, amplitude, and period) are also estimated and attributed to this detection. All other detections in the studied interval are discarded and the next detection has to be beyond 4 s from the found one.

It is often observed that the P-wave coda may contain signals similar to the direct P-wave (e.g. free surface reflections). Because of this observation we also prohibit any other arrival within 4 s despite the *CC* may exceed the threshold. This interval might be extended but then the risk to miss a valid signal from a different event also rises.

Before the start of event building, all arrivals have to pass a number of quality checks including the estimation of azimuth and slowness using the *f-k* analysis of the cross correlation traces. However, the possibility cannot be excluded that some higher correlation coefficients may be related to strong signals from sources far away from the master event or associated with coherent noise. To validate the signals detected on the *CC* traces we use *F*-statistics. Specifically, we calculate maximum $F_{prob}$ for all detections in the time window 4 s from their onset times. For that, *F*-statistics in a running 2 s window is estimated using the approach developed by Douze and Laster (1979). In this study, we tentatively put $F_{prob}>0.3$ as the threshold for a valid signal.

Figure 6 depicts the number of qualified (*CC*, $SNR_{CC}$, SNR, $F_{prob}$, azimuth and slowness residuals) arrivals as a function of *CC* threshold at 14 IMS array stations. These arrivals were found by cross correlation using all 141 master events. Since the master events have the number of stations from 3 to 14 the total number of arrivals depends on the frequency of participation of a given station in the master events. Table 5 lists the participation frequency for all 14 stations with the largest input from MKAR – 136 master events. ZALV participates in 130 masters and PETK only in 11. Altogether there are 1025 station (P and Pn) patterns for 141 events, i.e. approximately 7 patterns per master on average.

When interpreting the total number of arrivals one has to take into account that different masters predominantly find the same signals as they are characterized by a good cross correlation. The total number of arrivals reflects all valid master/slave pairs with different correlation coefficients and relative positions.

**Table 5. Station participation in the master events**

| sta | AKASG | ARCES | CMAR | FINES | GERES | KSRS | MJAR | MKAR | NOA | PETK | SONM | WRA | YKA | ZALV |
|---|---|---|---|---|---|---|---|---|---|---|---|---|---|---|
| # | 54 | 53 | 73 | 107 | 56 | 21 | 27 | 136 | 70 | 11 | 117 | 57 | 113 | 130 |

The largest total numbers belong to MKAR, which is the closest station. With the *CC* $CC_{tr}$ increasing from 0.15 to 0.40 the number of arrivals at MKAR drops from 105438 to 27377. In the lin-log scale in Figure 6, after some corner $CC_{tr}$, this fall can be approximated by linear line that manifests an exponential distribution. The transition from a constant level to an exponent fall likely manifests the change from noise detection to true signals from real events and exponential decay of cross correlation with distance.

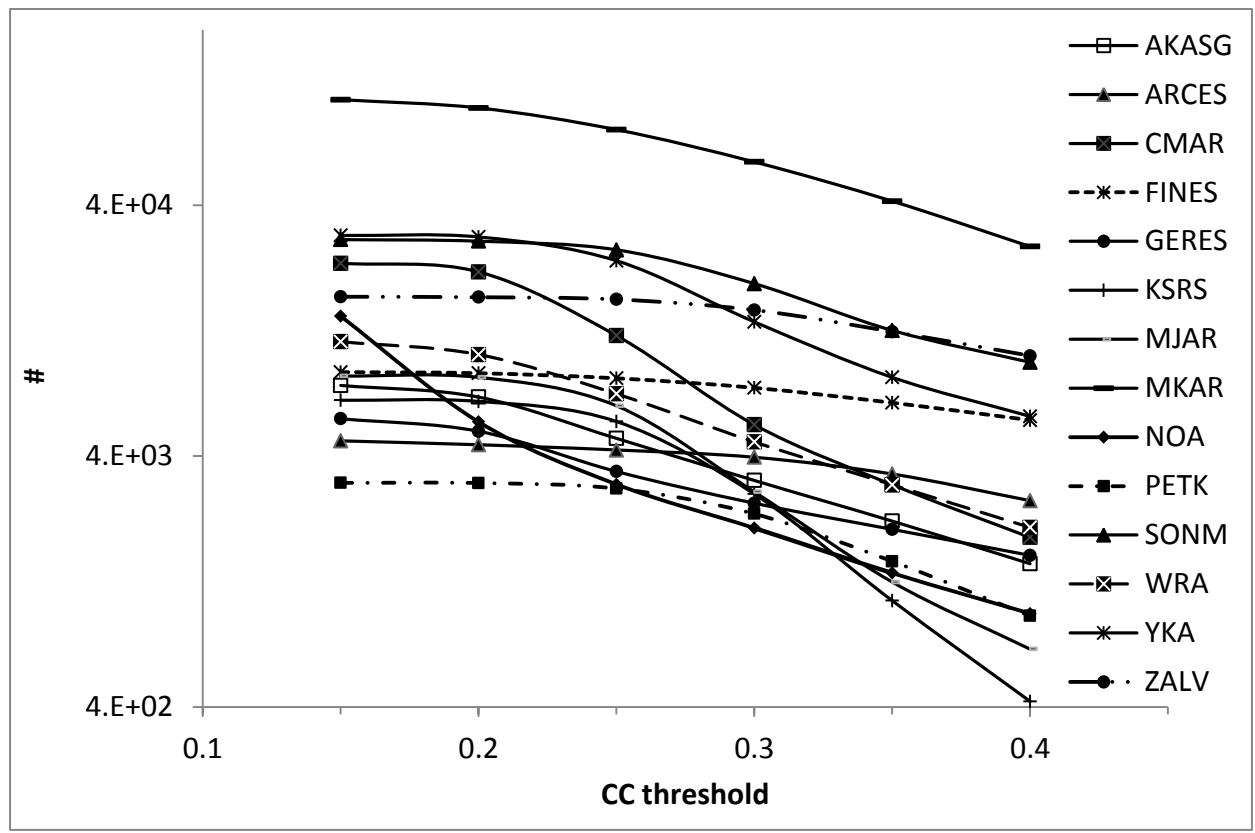

**Figure 6. The number of arrivals found by cross correlation at 14 IMS stations from 141 master events as a function of *CC* threshold.**

Form Figure 6, the corner $CC_{tr}$ for MKAR is between 0.15 and 0.25. For SONM and YKA, the corner *CC* is slightly above 0.25. For CMAR, one can observe a linear roll-off from 0.2. FINES and ZALV are likely characterized by a corner $CC_{tr}$ of 0.3. NOA demonstrates an unusual behavior – a convex (downward) curve with the quasi-linear portion from $CC_{tr} = 0.3$. This observation supports our earlier assumption on the dependence of $CC_{tr}$ on source region and station. It also gives an accurate tool to separate the flux of true arrivals at a given station from noise detections with a predefined rate of false alarms.

Our first task is to find as many REB events as possible using these cross correlation detections. The success of this task depends on the definition of a found phase and a found event. We follow the approach adopted at the IDC and consider an REB phase a found one when there is at least one cross correlation arrival within ±4 s interval. In IDC interactive processing, an analyst can retime P-wave arrival within this interval which might be considered as the uncertainty of arrival picking. Apparently, one needs at least three REB arrivals to be found with a given master event to consider the REB event as found by cross correlation.

Figure 7 summarizes principal results of the search for REB events as expressed by the cumulative cross correlation coefficient, $\Sigma CCj$, calculated for all found phases for a given master/slave pair. The results are arranged as a 142 by 142 matrix with the master events ordered in time from top to bottom. The cumulative cross correlation coefficients for the main shock as a master create the first line in the matrix. Since CC exceeds 0.2 at three or more station, the $\Sigma CCj$ scale starts from 0.6. When $\Sigma CCj >3.0$, the found REB event is at the level of autocorrelation for a three-station event. Therefore, we limit the cumulative $CC$ scale by 3.0. The $m_b$ scale shows (IDC) magnitudes of the master events in the range between 2.0 and 6.0. The $nsta$ scale shows the number of stations in the master event in the range from 3 to 14. Several masters have magnitude above 4.0 but a few stations. Three events with $m_b$(IDC)~4.3 include all 14 stations and one event with $m_b$(IDC)=3.86 includes 12 stations. These are candidates for the final master event set covering the area around the main shock in routine XSEL building.

Auto-correlation (diagonal cells) finds all REB events except the one with two templates. However we are interested in those REB events which are found by other masters. Figure 8 shows the number of found REB events as a function of magnitude, as obtained from Figure 7. The largest number of found REB by one master is 113 and belongs to the event with $m_b$(IDC)=5.05, i.e. to the largest aftershock. This master also finds the largest number of events overall – 165, which includes 113 REB events. The second best master with $m_b$(IDC)=4.52 finds 106 REB events (137 in total), and the master with $m_b$(IDC)=4.28 finds 152 events in total (97 REB).

The number of REB events found by a given master is almost constant for magnitudes between 4.1 and 5.05. Below 4.0, this number suffers a near-linear decay with magnitude. From five events with magnitude below 3.0 three found only themselves by auto-correlation. Other three events with auto-correlation only have $m_b$(IDC) of 3.16, 3.49 and 4.26(!). In total, only the main shock and 134 aftershocks can be used as master events. There are also 5 REB masters which found only one REB event in addition to auto-correlation.

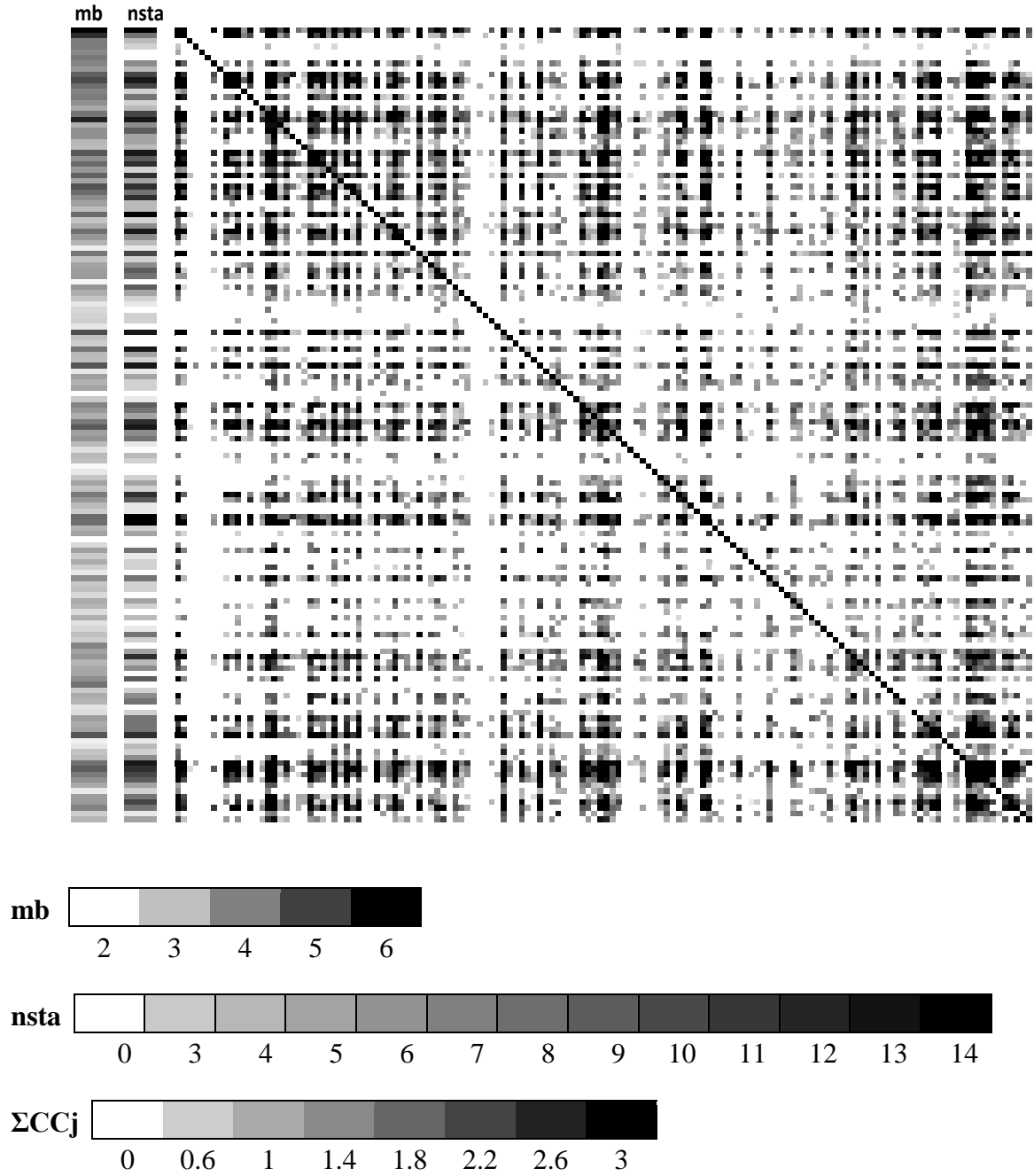

**Figure 7. Cumulative cross correlation coefficients, ΣCCj, for all pairs of 142 REB events with at least three stations having cross correlation coefficient above 0.2. Master events are ordered in time from top to bottom: cross correlations with the main shock create the first line in the matrix. The $m_b$ scale shows magnitudes of the master events. The nsta scale shows the number of stations in the master template. Some masters have large magnitude but few stations.**

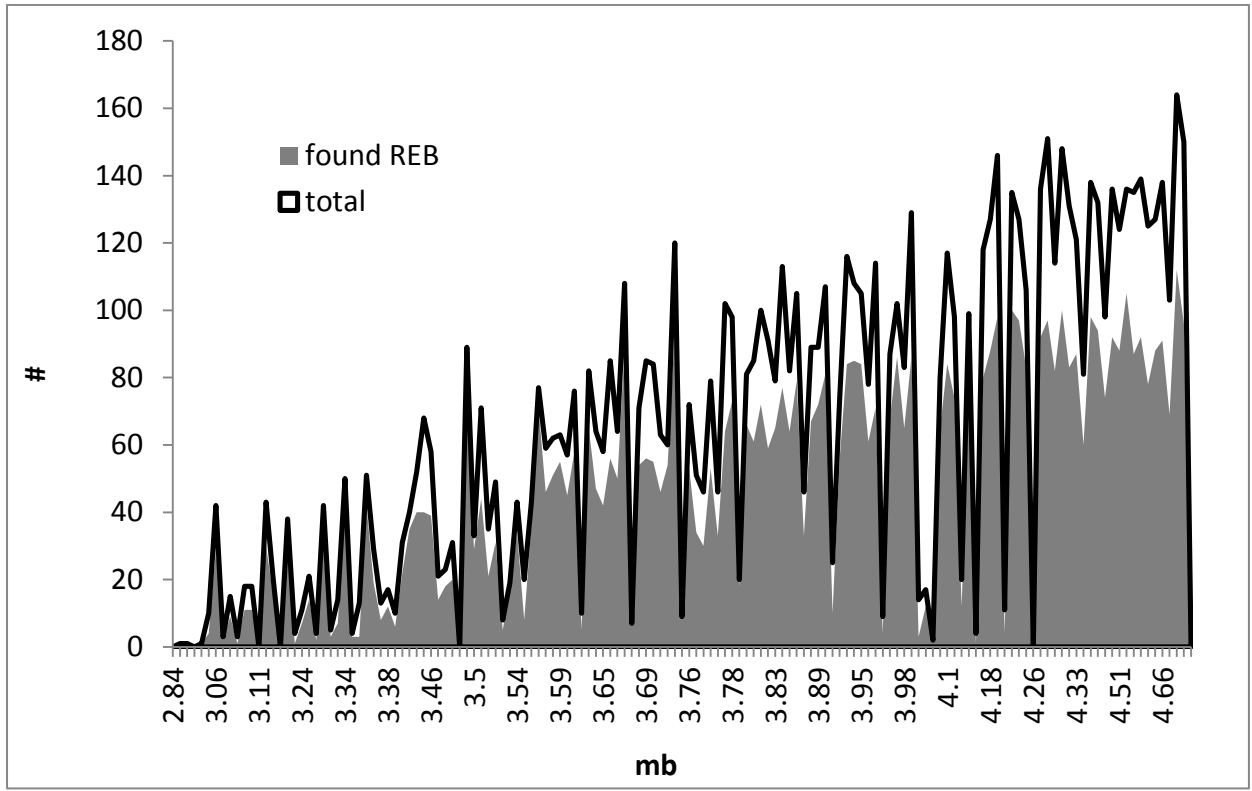

**Figure 8. The number of REB events and total number of events found by selected master events, ordered by magnitude. The largest total number of found events (165) and REB events (113) corresponds to the same event with $m_b$=5.05.**

Another view on Figure 7 reveals the number of REB events which were not found by any master. There are nine not found (by cross correlation) aftershocks with only two from the set of worthless masters. There are also 5 aftershocks which were found by only one master. Altogether we have 7 worthless masters, 7 not found events and 5 poorly defined events, which found or were found by only one event.

There are two principal reasons for an arrival in the REB not to be found by cross correlation with other master events. There is a valid arrival detected by cross correlation but it is out of the predefined 4 s widow; such an REB arrival is likely a later phase. The not found REB arrival is a misassociated phase from a different event and thus fails to cross correlated with any of correct templates. In both cases, the relevant REB events do not contradict IDC rules and guidelines because all defining parameters are within the uncertainty bounds adopted for travel time, azimuth and slowness residuals.

The presence of somehow misassociated phases in the REB does not preclude cross correlation from finding valid arrivals where they have to be. For some of the poorly built aftershocks, cross correlation has detected three and more phases when only one or two REB arrival was matched by time. Even for those aftershocks which are found by many master events additional arrivals have been found by cross correlation. These additional arrivals increase the cumulative *CC*s in Figure 7.

Figure 9 presents a matrix with the cumulative *CC* obtained for all cross correlation arrivals (in and not in the REB). There is a tangible improvement in event finding – eight REB events have one or two phases shared with at least one event built with cross correlation by three or more primary stations. In other words, we would build more reliable REB events if cross correlation detections were used.

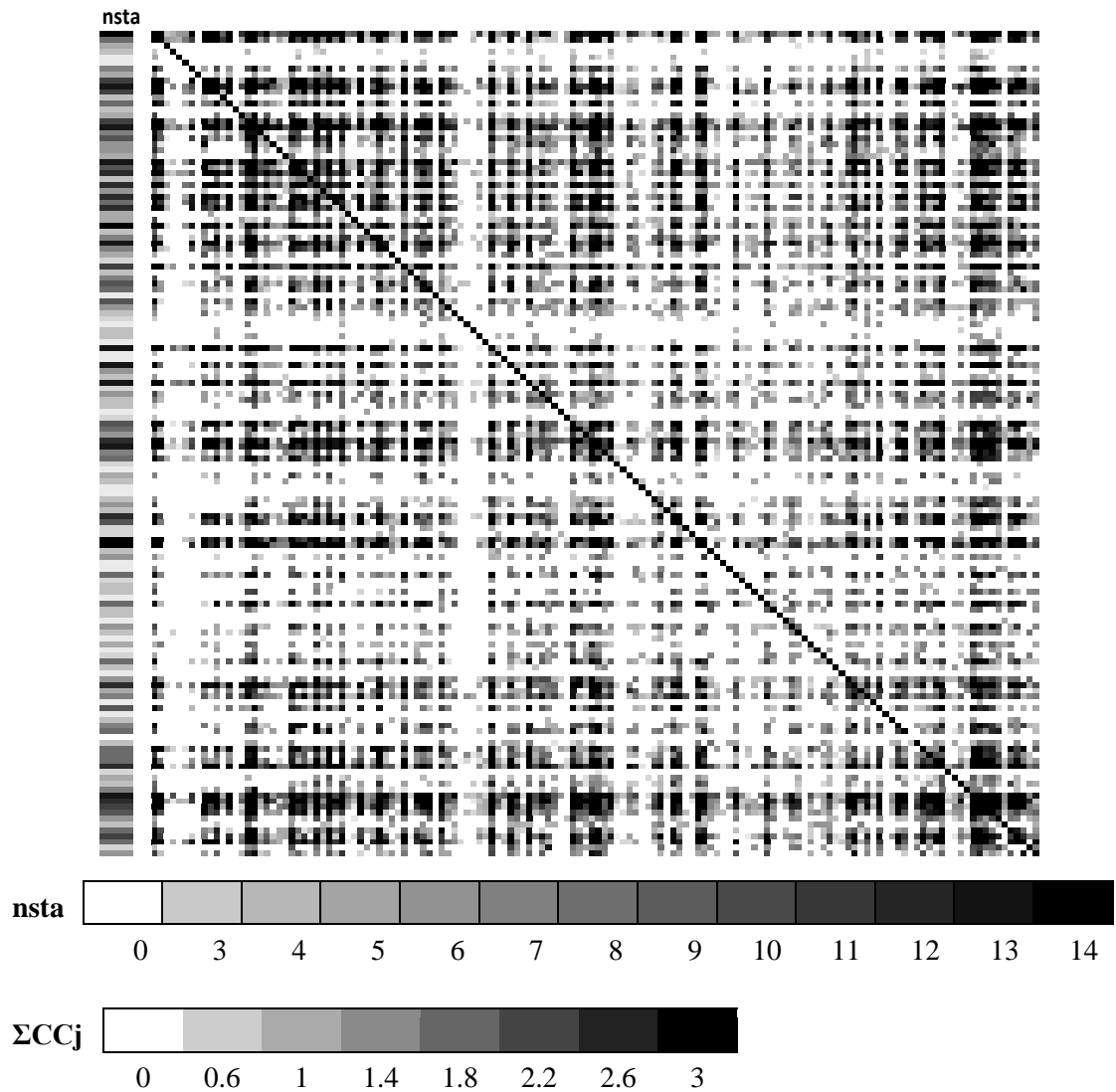

**Figure 9. Cumulative cross correlation coefficients, *ΣCC*j, for all stations with *CC*s above 0.2. Master events are ordered in time from top to bottom.**

There are many questions left on the relations between the REB and cross correlation arrivals in the studied aftershocks. However, the principal results of the cross correlation study obtained so far allow addressing the first and second tasks formulated in the beginning of this Section. We have found a larger part of the aftershocks using other aftershocks as master events. One hundred seven aftershocks have been found by at least 10 master events and one hundred ten masters have found at least 10 aftershocks. Therefore, approximately 110 from 140 (142 less the main shock and the event with two stations) REB events (79 per cent) can be considered as mutually consistent ones.

Among the residual 30 REB events one may distinguish 13 less reliable master events with 4 to 9 found aftershocks and 17 poor events which, in practice, were able to find only themselves. These 17 REB events have to be considered for internal consistency. Figure 9 gives an obvious hint that many of these 30 events could be built in a more reliable manner when *CC* is used. In any case, none of these 30 REB events should be used as master ones. The pre-selection of master events is the second part of task 2.

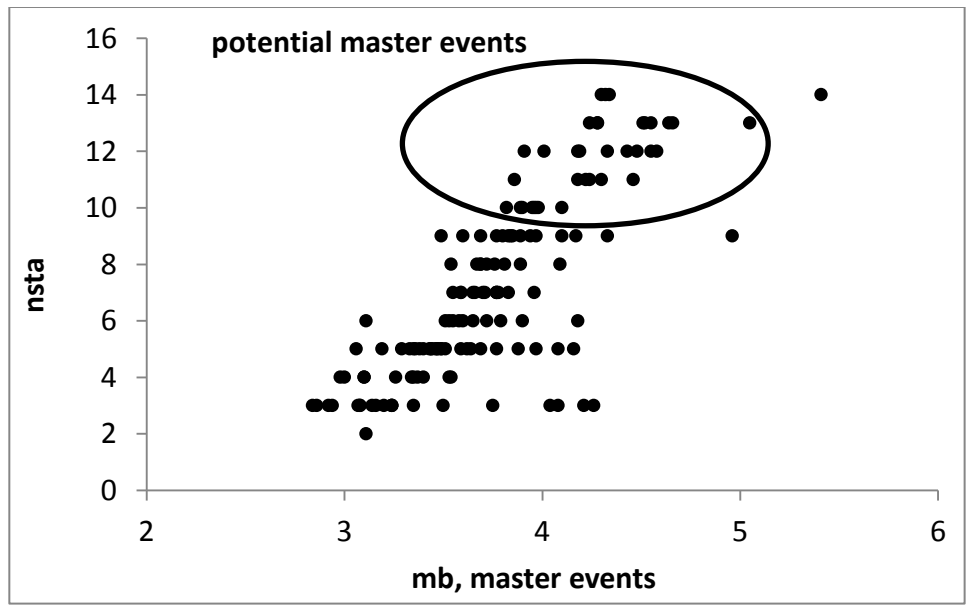

**Figure 10. Distrubution of the number of (primary array) stations in master events depending on magnitude.**

Figure 10 shows the number of templates in all aftershocks. Potential master events have to have the largest possible number of templates, i.e. ten and more. This guarantees the best coverage in presence of varying seismic noise at IMS stations. When noise is high at several stations one still has a chance to find enough signals to build an XSEL event. Potential master events should also be able to find as many aftershocks as possible, as shown in Figure 8. Now we have to intoduce the procedure of XSEL events building.

We start with one master event. All *CC* and $SNR_{CC}$ values are already estimated for all four different frequency band templates for all stations in the master event. After all conflicts between arrivals in various frequency bands are resolved (see Section 2) one has a set of detections with their arrival times for each station, $t_{ij}$, where *i* is the index of the *i*th arrival at station *j*. Apparently, all valid arrivals should belong to some events in proximity to the master event. These are the events we are seeking for. The travel times from these sought events to the relevant stations can be accurately approximated by the master/station travel times, $tt_j$. Using the approximated travel times (same for all events around the master one) and the measured arrival times one can calculate origin times for all detections:

$$ot_{ji} = t_{ij} - tt_j$$

As a result, one has a set of origin times at several stations instead of arrival times. Origin time is a natural characteristic of source and when three or more stations detect appropriate signals within a few second one can associate these arrivals with a unique event. (Here we follow up the IDC event definition criteria.) At the initial stage, we average all associated origin times and assign the estimated value to the event origin time. One can also use the median value or various weighted sums. There is no need to locate the built event globally since it has to be close to the master one. When necessary, relative location can be carried out by a double difference algorithm.

Overall, this process can be called "local association", *LA*, in line with the name of global association, GA. Indeed, only phases from local events should be associated. The *LA* does not see any events beyond the radius of correlation. That is why two DPRK master events did not find any from hundreds events occurred globally during the five days after the second test. The *LA* had no local arrivals to associate.

We have started with a six-second time window in the *LA* for the association into a single event. Considering the travel time uncertainties in the GA (dozens of seconds), this is a very short interval for origin times. It corresponds to the difference in travel times between the master event and an event on the rim of cross correlation zone (say, 50 km). For a P-wave with 0.05 s/km slowness the travel time difference is of 2.5 s. For two stations in opposite directions, two 2.5 s travel time residuals give a 5 s difference in origin time. This is the worst case scenario. A 1 s uncertainty in onset time may add 2 s to the origin time difference. We have tested longer windows and found six seconds to provide almost as many events as a nine second window. For the latter window, more noise phases might be wrongly associated and additional events are likely less reliable.

At regional distances, Pn-waves have larger slowness but cross correlation coefficient decays much faster with spacing between events. This is the effect of the highly inhomogeneous crustal and upper mantle structure, where the Pn-waves propagate. As a result, the association window of six seconds still works efficiently.

There is an important enhancement of the *LA* process based on relative magnitudes of associated arrivals. As shown in Section 2, all arrivals associated with a given event should have *RM* estimates within some predefined bounds separating the genuine and bogus signals. We have adopted a tentative value of 0.7, which is much smaller than a similar magnitude difference of 2.0 used in IDC automatic processing. This threshold is then tested on the full set of found events.

The *LA* is a simplistic process compared to the global association. A big advantage of the *LA* is a reduced flux of arrivals at a given station from a given master event. At the same time, the total number of arrivals at a station may grow relative to that from the current IDC detector. We have already mentioned that the cross correlation detector can find valid signals with (standard) SNR<0.7 which are not seen by the current IDC detector. The increased number of arrivals can be effectively split into a large number of independent sets associated with different masters.

When several master events are close in space, like in the studied aftershock sequence, their templates may correlate with waveforms from the same event and thus create similar new events. To select one from a set of similar events with close origin times we first count for all individual masters the number of stations used in the *LA*. When several master events have the largest number of stations we select the one with the highest cumulative cross correlation coefficient, $\Sigma CC_j$. By definition, this event is the most reliable and its parameters are written into the database. All other events in the set are rejected. Thus, for a multiple set of master events the *LA* provides a unique set of found events.

We have tested the influence of the number of master events on the final catalog of new events. When a found event has origin time within ±15 s from any of the REB events (reliable and poor ones) it is considered as associated with the REB not as a new one. We are looking for distinct new events which cannot be confused with the REB events. To populate the set of master events we progressively included more and more masters and calculate the number of new events. To begin with, we used the best master event which has previously found 112 REB events and 165 events in total. Then we added the second and third best master events from Figures 8 and 10. Two sets of 10 and 27 (all aftershocks with 10 and more stations) events have also been tested for the total number of new events. The latter number includes.

Figure 11 depicts the number of new events as a function of the master set size for three different *CC* thresholds: 0.15, 0.20, 0.25 and 0.35. The $CC_{tr}$ defines the number and quality of used arrivals. At some IMS stations the threshold can be as low as 0.20. Other stations need 0.3 and even larger threshold in order to reduce the portion of noise arrivals.

The number of new events grows with the size of master event set. For $CC_{tr}$=0.20, there are 36 new events for the best master event. For 27 masters, the number of new events is 94. This is a natural trend - since the cross correlation coefficient falls exponentially with distance a denser master grid should find more events by virtue of proximity. Ultimately, one might design an iterative procedure to find all possible events (Harris and Dodge, 2011). First, all (reliable) REB events are used as masters to find new events. After an interactive review, these new events are included in the REB and then used as masters to find the next portion of new events. The process is repeated before there is no new event. To enhance this procedure one may combine best templates from different but well correlated events. For example, one may use as templates stations YKA and MKAR from event 1 and stations ARCES and FINES from event 3 to create a synthetic event. Moreover, it is possible to combine waveforms from different stations to create synthetic waveforms, as proposed by Harris and Paik (2006).

For an area covered by seismic events spaced by less than, say, 30 km (continuos coverage), one needs only one reliable event to start this iterative procedure which will end up in a complete XSEL (or REB) bulletin. The REB events distribution shown in Figure 1 assumes that the IDC likely needs only one or a few events to start the process which will find all reliable

REB events and reorder them according to their cross correaltions. This reordering also means accurate relocation with the best located events (e.g. the events with ground truth coordinates) defining the absolute locations and confidence ellipses for less succesful "cross correlation neighbors". As we have demonstrated, there exist unreliable (in cross correaltion sense) REB events. A good portion of these events could be re-built with cross correlation arrivals replacing the misassociated phases. However, a few REB events cannot be healed by cross correlation. In addition to the recovered and quality checked REB, one will obtain a set of new reliable REB events which might be as large as the REB itself. For the number of existing REB events and the total length of waveforms obtained since the launch of IDC in 2001, this iterative process needs a supercomputer.

Higher *CC* thresholds reduce the final new event bulletin (i.e the XSEL less REB) and likely improve its quality. For $CC_{tr}$= 0.35, there are only 12 new events for the best master and 32 new events for 27 masters. Altogether, for all 141 master events there are 9707 new events built (69 new events per a master event) for $CC_{tr}$=0.15 and only 5984 for $CC_{tr}$=0.35 (42 per a master event). Cross correlation coefficient of 0.35 is not often among the arrivals in the XSEL and likely belongs to the closer events with higher magnitudes.

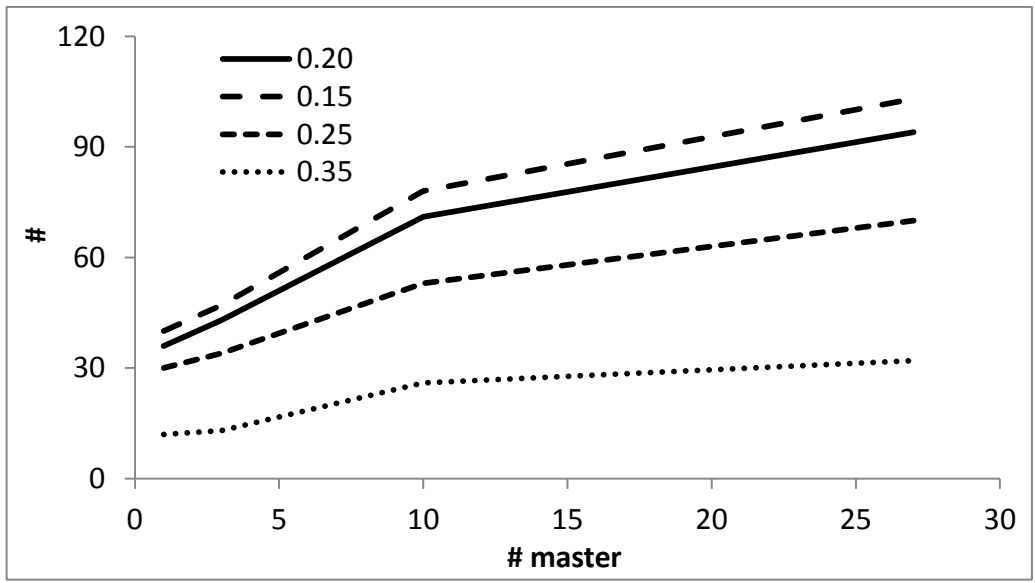

**Figure 11. The number of new events found by various sets of master events for different CC thresholds: 0.15, 0.20, 0.25, 0.35.**

A crucial part of the third task is interactive checking of all new events. We have to draw a distinct line between reliable and poor new events. Only a human can make the final decision on event quality. We started with $CC_{tr}$=0.2 and the set of 10 master events. This set finds 126 from 141 REB events by cross correlation at three and more stations. The residual 15 events are likely internally inconsistent. There are 71 new events to be reviewed and those with a larger number of reliable stations are of a higher priority.

The station reliability can be estimated from Table 6 where the portion of all arrivals obtained by *CC* used in the final XSEL is listed. For $CC_{tr}$=0.2, station NOA has the highest rate of ~57% of all arrivals used in the XSEL (126 REB and 71 new events). AKASG, FINES, WRA and GERES are characterized by rates above 30%. At the same time, the input of PETK and KSRS is negligible. These stations are likely able to detect signals from the events with magnitude above 5.0.

**Table 6. The percentage of cross correlation arrivals used in the XSEL**

|       | 0.15 | 0.2  | 0.25 | 0.35 |
|-------|------|------|------|------|
| AKASG | 47.5 | 51.4 | 66.1 | 81.2 |
| ARCES | 26.4 | 26.9 | 27.3 | 31.0 |
| CMAR  | 13.4 | 14.3 | 23.0 | 52.2 |
| FINES | 40.9 | 40.8 | 40.8 | 40.8 |
| GERES | 27.6 | 30.7 | 42.0 | 60.1 |
| KSRS  | 1.8  | 1.8  | 1.9  | 5.4  |
| MJAR  | 10.3 | 10.3 | 12.5 | 43.6 |
| MKAR  | 7.2  | 7.6  | 8.4  | 10.7 |
| NOA   | 24.6 | 56.6 | 76.6 | 82.9 |
| PETK  | 1.9  | 1.9  | 1.9  | 2.8  |
| SONM  | 19.4 | 19.4 | 19.8 | 30.7 |
| WRA   | 34.3 | 37.4 | 44.1 | 57.9 |
| YKA   | 22.4 | 22.4 | 25.8 | 49.2 |
| ZALV  | 30.1 | 29.9 | 29.1 | 30.9 |

An important exclusion from the reliability hierarchy is MKAR. It generates many arrivals but 90 per cent of them is not used in the XSEL. This is a distinct signature of low reliability. However, the absence in the XSEL does not mean that an arrival is worthless or bogus. The Gutenberg-Richter law implies that there should be more events with smaller magnitudes which can be detected by two stations or even by one station only. As the closest regional array, MKAR has to find valid signals from these smaller events which cannot meet the IDC event definition criteria. This consideration is also applicable to those arrivals at reliable stations which are not in the XSEL. They may also create many two-station events which we do not review.

**Table 7. Number of new events with a given number of stations for various sets of master events**

| # masters | 1 | 3 | 10 | 27 | 141 |
|---|---|---|---|---|---|
| nsta=3 |  | 20 | 25 | 46 | 65 | 89 |
| nsta=4 |  | 9 | 10 | 12 | 14 | 19 |
| nsta=5 |  | 6 | 7 | 12 | 10 | 10 |
| nsta=6 |  | 1 | 1 | 1 | 4 | 4 |
| nsta=7 |  | 0 | 0 | 0 | 1 | 1 |
| total |  | 36 | 43 | 71 | 94 | 123 |

There is one event with $m_b$(IDC)=3.9 which occurred several minutes after the main shock when the level of microseismic noise was elevated by secondary phases of the main shock and big aftershocks. This event included only 4 stations. The biggest event has $m_b$(IDC)=4.2 and occurred two hours after the main shock. The smaller new event has $m_b$(IDC)=2.8.

Figure 12 displays original waveforms and the IDC solution for a smaller event with $m_b$(IDC)=3.4 which includes 7 stations with one auxiliary array BVAR. All signals are clear and reliable. Further work is needed to review other events from the set of 71, and then from the set of 94 as obtained by 27 master events, but even the initial effort has shown a large number of events missed in the REB and a high success rate of the cross correlation pipeline. Definitely, there should be some bogus events among those 71 or 94 built for $CC_{tr}$=0.2. One has to decide on the tolerable rate of false alarms and to tune all defining parameters accordingly.

Interactive analysis is a time consuming procedure and an experienced lead analyst has reviewed with IDC rules and guidelines only 40 from 71 events. The analyst started with the events with the largest number of stations (see Table 7) but also reviewed several less reliable cases. There were built 37 new REB events (success rate 93%) and their reviewed IDC solutions are listed in Table 8 with locations, origin times and magnitudes. One of the reviewed events was actually an REB event with $m_b$(IDC)=4.4 approximately 2000 km off the main shock. However, it was big enough to generate quasi-sinusoid waveforms of twenty and more seconds at teleseismic distances. It was wrongly built by cross correlation as an aftershock because all three primary IMS stations which built this event were close (NOA, FINES and ARCES) and at the same great circle with the main shock and the REB event. This problem was fixed by checking the azimuth gap and relative magnitude.

**Table 8. IDC solutions for 37 new events reviewed by analysts**

| lat | lon | Day | Time | nass | ndef | mb | ML |
|---|---|---|---|---|---|---|---|
| 35.48 | 80.56 | 20/03/2008 | 22:45:32 | 4 | 4 | 3.9 | 3.5 |
| 35.61 | 81.55 | 20/03/2008 | 23:28:20 | 6 | 6 | 3.4 | 3.0 |
| 34.77 | 81.11 | 20/03/2008 | 23:57:10 | 7 | 7 | 3.8 | 2.8 |
| 36.46 | 81.96 | 21/03/2008 | 00:08:12 | 3 | 3 | 2.9 | 2.7 |
| 34.88 | 80.92 | 21/03/2008 | 00:12:07 | 4 | 4 | 4.2 | - |
| 34.08 | 81.28 | 21/03/2008 | 00:13:02 | 4 | 4 | 4.1 | - |
| 34.91 | 81.23 | 21/03/2008 | 00:15:45 | 5 | 5 | 3.6 | 3.0 |
| 35.31 | 81.33 | 21/03/2008 | 00:41:10 | 5 | 5 | 3.6 | 3.1 |
| 35.47 | 81.08 | 21/03/2008 | 01:07:55 | 7 | 7 | 3.4 | 3.3 |
| 35.35 | 81.46 | 21/03/2008 | 01:15:43 | 4 | 4 | 3.7 | - |
| 34.91 | 82.30 | 21/03/2008 | 01:43:01 | 7 | 6 | 3.4 | 3.0 |
| 35.20 | 80.50 | 21/03/2008 | 02:35:33 | 6 | 5 | 3.1 | 2.9 |
| 35.21 | 81.31 | 21/03/2008 | 02:53:49 | 4 | 4 | 3.4 | 2.2 |
| 35.04 | 80.88 | 21/03/2008 | 04:33:13 | 4 | 4 | 3.2 | 3.1 |
| 35.49 | 81.19 | 21/03/2008 | 06:25:59 | 6 | 6 | 3.4 | 3.5 |
| 35.75 | 81.20 | 21/03/2008 | 07:30:36 | 4 | 4 | 3.1 | 2.8 |
| 35.26 | 80.82 | 21/03/2008 | 07:33:52 | 7 | 7 | 3.4 | 2.5 |
| 35.21 | 80.88 | 21/03/2008 | 07:37:01 | 6 | 6 | 3.1 | 2.9 |
| 35.23 | 81.64 | 21/03/2008 | 08:21:26 | 4 | 4 | 3.0 | 3.0 |
| 35.84 | 80.71 | 21/03/2008 | 09:31:23 | 4 | 4 | 3.0 | 2.9 |
| 35.32 | 81.22 | 21/03/2008 | 09:45:47 | 5 | 5 | 3.4 | 3.0 |
| 35.21 | 80.84 | 21/03/2008 | 09:57:07 | 4 | 4 | 3.1 | 2.6 |
| 35.07 | 82.47 | 21/03/2008 | 11:17:15 | 8 | 8 | 3.4 | 3.0 |
| 35.72 | 81.21 | 21/03/2008 | 13:29:53 | 4 | 4 | 2.8 | 3.2 |
| 35.22 | 81.18 | 21/03/2008 | 14:11:55 | 9 | 9 | 3.4 | 3.2 |
| 36.35 | 81.78 | 21/03/2008 | 21:13:34 | 7 | 6 | 2.9 | 3.2 |
| 36.02 | 81.69 | 21/03/2008 | 21:37:47 | 6 | 5 | 2.8 | 2.9 |
| 37.01 | 80.18 | 21/03/2008 | 21:57:17 | 3 | 3 | 3.2 | 2.9 |
| 35.30 | 81.10 | 21/03/2008 | 22:43:02 | 7 | 7 | 3.5 | 2.8 |
| 35.39 | 80.65 | 22/03/2008 | 07:37:56 | 6 | 6 | 3.2 | 2.8 |
| 35.61 | 81.20 | 22/03/2008 | 21:53:18 | 7 | 7 | 3.5 | 2.8 |
| 35.63 | 81.77 | 23/03/2008 | 02:08:54 | 11 | 9 | 3.4 | 3.1 |
| 35.85 | 80.71 | 23/03/2008 | 10:36:31 | 5 | 5 | 3.0 | 2.9 |
| 35.37 | 81.29 | 23/03/2008 | 10:46:12 | 6 | 6 | 3.4 | 2.8 |
| 35.43 | 81.12 | 24/03/2008 | 00:03:08 | 4 | 4 | 3.5 | 2.8 |
| 35.21 | 80.70 | 24/03/2008 | 09:38:48 | 7 | 7 | 3.8 | 3.4 |
| 35.75 | 80.99 | 24/03/2008 | 17:52:46 | 5 | 4 | 3.0 | 2.9 |

Signals for the residual two new events were not found by standard methods of interactive analysis. This situation was expected because cross correlation is definitely able to find valid signals below the noise level. Therefore, the negative result of interactive review does not imply that these events were wrongly built by cross correlation and the success rate might actually be higher than 93 %.

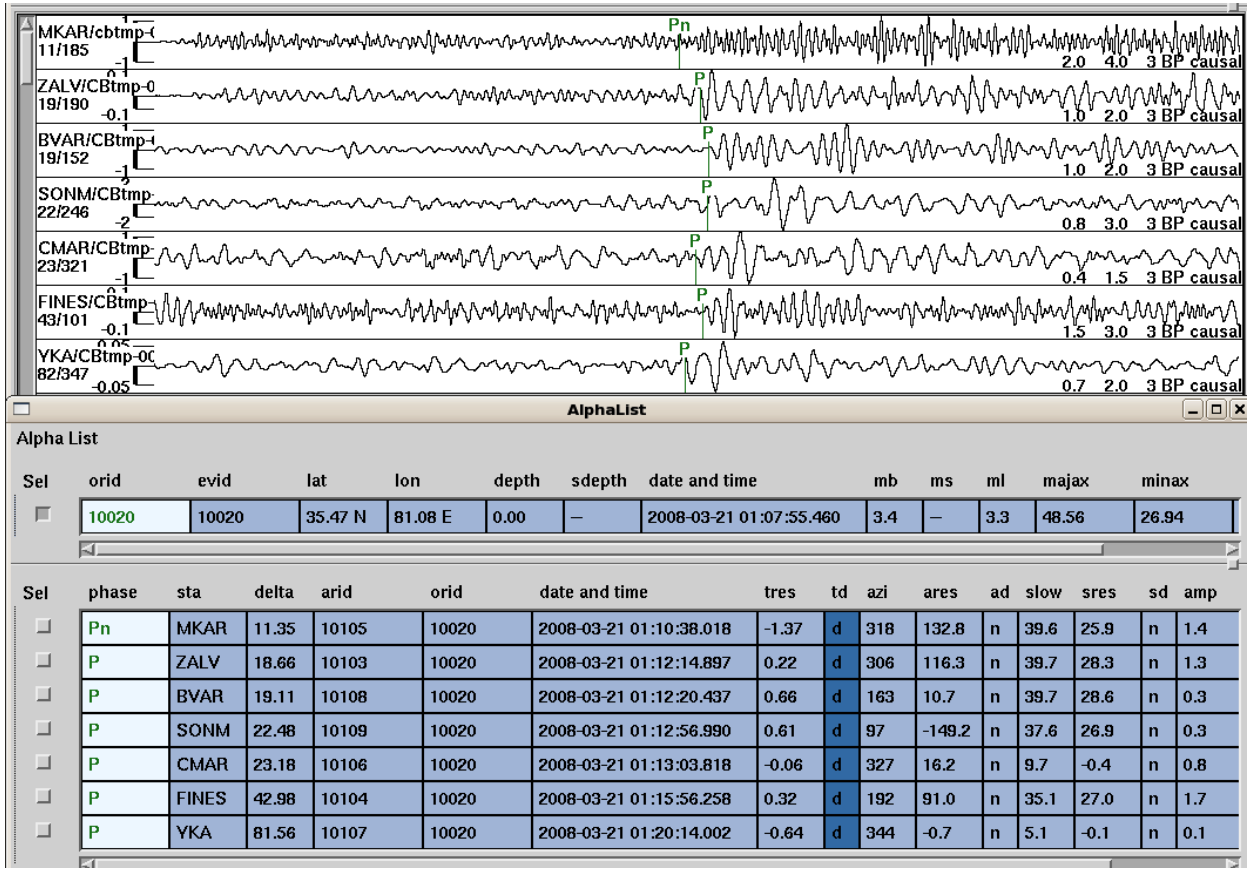

**Figure 12. An example of new event built using detections obtained from cross correlation coefficients at six (!) primary stations with $m_b$(IDC)=3.4. The event definition criteria are all matched. BVAR is an auxiliary array which was not used as a template but found in interactive analysis. Origin time: 21-03-2008, 01:07:55.46; location: 35.47N, 81.08E. Seven waveforms (beams) were filtered with different filters shown in the Figure. The list of stations shows the onset times picked by analyst. Azimuths and slownesses for these detections were estimated by automatic procedures from the original waveforms and demonstrate high residuals.**

In many cases, the number of associated phases, *nass*, and the number of time defining phases, *ndef*, in Table 8 is larger than the number of stations in the relevant XSEL events. This difference has two sources. The analyst may add arrivals detected by primary 3-C stations or by auxiliary stations; both types are not used for cross correlation. Secondary arrivals may also be added at all station type. For example, the closest auxiliary station AAK (see Figure 5) reports Pn- and Lg- phases which are used to constrain IDC locations and confidence ellipses. Auxiliary station BVAR is actually a very sensitive array which also detected a good portion of the aftershocks.

All new events in Table 8 were reviewed and thus located using standard IDC software with the parameters of arrivals obtained in standard automatic processing. In many cross correlation studies a double-difference (DD) algorithm is used. It is based on the relative travel times from the master and slave event (e.g. Schaff *et al.*, 2004). The DD method allows for an accurate location of the slave event relative to the master event what is of importance for many seismological and tectonic applications. For the IDC purposes, absolute locations are of the highest priority. When a master event is mislocated by tens of kilometers (usual values for events with a few stations) the relative location introduces systematic errors in all cross correlated events. An independent location of many events with travel time, azimuth and slowness residuals distributed approximately normally is likely able to provide a more reliable solution on average (Kitov and Koch, 2007). Therefore, we will use the standard IDC location procedure in interactive review before the set of master events is not perfectly located in absolute terms.

**Table 9. Cumulative share of arrivals associated with at least one event below given CC levels**

|       | 0.2  | 0.25 | 0.3  | 0.35 | 0.4  | 0.45 | 0.5  |
|-------|------|------|------|------|------|------|------|
| AKASG | 0.01 | 0.13 | 0.32 | 0.51 | 0.66 | 0.77 | 0.84 |
| ARCES | 0.00 | 0.03 | 0.05 | 0.11 | 0.25 | 0.42 | 0.55 |
| CMAR  | 0.00 | 0.11 | 0.31 | 0.49 | 0.65 | 0.76 | 0.84 |
| FINES | 0.00 | 0.01 | 0.04 | 0.10 | 0.18 | 0.28 | 0.38 |
| GERES | 0.00 | 0.05 | 0.11 | 0.20 | 0.31 | 0.41 | 0.52 |
| KSRS  | 0.02 | 0.12 | 0.31 | 0.50 | 0.60 | 0.64 | 0.74 |
| MJAR  | 0.00 | 0.07 | 0.20 | 0.37 | 0.53 | 0.68 | 0.79 |
| MKAR  | 0.00 | 0.04 | 0.12 | 0.23 | 0.38 | 0.55 | 0.71 |
| NOA   | 0.02 | 0.26 | 0.48 | 0.64 | 0.74 | 0.82 | 0.88 |
| SONM  | 0.00 | 0.01 | 0.05 | 0.11 | 0.18 | 0.25 | 0.33 |
| WRA   | 0.01 | 0.20 | 0.39 | 0.54 | 0.66 | 0.75 | 0.81 |
| YKA   | 0.00 | 0.05 | 0.17 | 0.33 | 0.48 | 0.60 | 0.70 |
| ZALV  | 0.00 | 0.00 | 0.01 | 0.02 | 0.07 | 0.14 | 0.25 |

The fourth task also demands considerable human resources. One has to estimate all defining parameters: $CC_{tr}$, thresholds for $SNR_{CC}$ and standard SNR together with the lengths of STA and LTA windows, frequency bands and lengths of templates for various seismic phases (e.g. P, Pg, Pn, PKP, Sn, Lg), parameters of *F*-statistic and $F_{prob}$, the width of the *LA* association window, the minimum time spacing between arrivals at one station and between events built by the same master event, etc.

The problem of optimal thresholds is a typical task for machine learning. Here, we report a few principal relationships based on 45585 arrivals associated with at least one event build by cross correlation with all 141 REB as master events. First is the probability for an arrival at a given station to be associated with an event as a function of *CC*. Table 9 lists the cumulative portion of arrivals associated with at least one event below given CC levels. For AKASG, a half of associated arrivals have *CC*>0.35. For ARCESS, only arrivals with *CC*>0.5 have the probability to be associated above 50 per cent. We excluded PETK from the list – it has only 27 associated arrivals, likely for the biggest events. It is worth noting that low *CC* does not preclude an arrival to be correctly associated with an XSEL event. Using Table 8 and Figure 6 one can define sound station-dependent thresholds of *CC*.

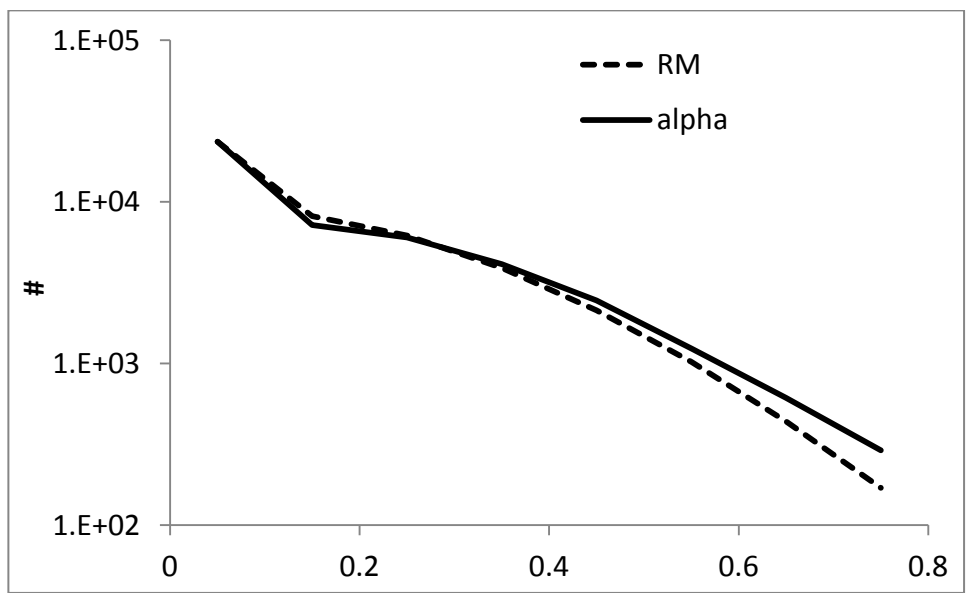

**Figure 13. Frequency distribution of *RM* and α residuals counted in 0.1-wide dimensionless unit bins as obtained from all XSEL events built by all 141 master events. In the lin-log scale, the number of residuals falls slightly faster than exponentially with only a few residual between 0.6 and 0.7. The residuals of relative magnitude have a smaller standard deviation 0.16 *vs.* 0.18 for α.**

Figure 13 depicts frequency distribution of *RM* and α residuals counted in 0.1-wide (dimensionless unit) bins as obtained from all XSEL events built by all 141 master events. For an event, the residuals are the absolute deviations of individual estimates from the average. We have already introduced a tentative *RM* threshold of 0.7. In the lin-log scale, the number of residuals falls slightly faster than exponentially with only a few residual between 0.6 and 0.7. This observation validates our preliminary assumption on the *RM* fluctuations for reliable events. One can also consider a smaller threshold. As expected, the set of *RM* residuals has smaller mean value (0.17) and standard deviation (0.14) compared to those for α: 0.19 and 0.16, respectively. Individually, the standard deviation varies in a narrow band between 0.16 for AKASG to 0.26 for WRA.

**Conclusion**

We have developed and tested a set of procedures allowing automatic detection and event building in accordance with the IDC event definition criteria for REB events. These procedures are arranged in a unique line, in parallel to the current IDC pipeline, and it is possible to issue an automatic XSEL (REB-ready) bulletin when all data from primary stations are available at the IDC. Therefore the XSEL is an analogue of SEL1 but with REB quality.

The core of detection and phase identification consists in cross correlation of incoming waveforms from primary array stations with a predefined set of waveform templates from accurately prepared master events. A number of standard and newly introduced parameters are estimated for all arrivals. Then the arrivals obtained from one master event at different stations are associated in XSEL events in the local association procedure. The event building procedure includes testing for dynamic consistency between arrivals as based on relative magnitudes.

The obtained XSEL events were compared to the existing REB and new events are reviewed by an experienced analyst. The results of cross correlation have validated the overall consistency of approximately 90% of the studied REB events. Approximately 10% of the REB events do not correlate well with these 90% and also with each other. They have to be reconsidered in order to reveal the reason of inconsistency. We have also added 37 new vents to the existing REB with dozens more event hypotheses to be reviewed.

All in all, we have demonstrated that several techniques based on cross correlation are able to significantly reduce the detection threshold of seismic sources in one specific region, and thus, worldwide and to improve the reliability of IDC arrivals and events by a more accurate estimation of defining parameters. The completeness of the REB could be significantly improved (and will be improved in the future) in automatic processing while fitting the event definition criteria. Moreover, there are several options on the way to the completeness: one can always balance the density of master events and computer recourses.

The rate of false alarms has also been reduced by several powerful filters compared to the proportion of rejected events in SEL3. The most efficient filter is cross correlation between waveforms which separates signals from a small area around a given master event and the whole outer space. Despite its higher sensitivity to seismic sources around a given master event, cross correlation trims away 99.9% of the overall arrival flux in routine IDC processing as irrelevant to local association. Only qualified arrivals are considered for association. The travel times from the master event to three or more primary stations allow constraining the origin times for new events in a narrow range of a few seconds. The *LA* process also includes relative magnitudes to resolve possible dynamic inconsistency between associated arrivals. The *RM* is a reliable characteristic of relative sizes with low scattering. Two additional filters, *f-k* analysis and *F*-statistics, are also based on correlation traces, demonstrate excellent performance and provide accurate estimates of azimuth and slowness. As a result, cross correlation may reduce the overall workload related to IDC interactive review and build a precise tool for quality check, for both arrivals and events.

Major improvements in automatic processing and interactive review achieved by cross correlation are illustrated by an aftershock sequence of a large continental earthquake. Exploring this event, we have described schematically further steps in the development of a processing pipeline parallel to the existing IDC one in order to improve the quality of the REB together with the reduction of detection threshold. The current IDC processing pipeline should be focused on the events in remote areas not properly covered by REB events and those events which include infrasound and/or hydroacoustic phases.

Our example shows that cross correlation can easily cope with automatic processing of mid-sized aftershock sequence, especially in new areas. Using only one big REB event found by automatic processing we are able to iteratively recover the whole sequence for the following five days. Moreover, we have found several REB events which should be reconsidered because they are believed to include misassociated seismic phases, which have no correlation with all master templates. Such arrivals would have been screened out in cross correlation processing. For the largest earthquakes, the area of aftershocks may cover tens thousands of square kilometers and the distance between remote aftershocks may reach several hundred kilometers. In this situation, the set of master events has to cover the entire area with a regular spacing. In some cases, there are no REB events in mid-size areas which are surrounded by seismically active regions. One may try to find new events using cross correlation and templates at the boarders of these blank areas and/or build synthetic templates by extrapolation of the existing templates. Fully synthetic templates are also an option with theoretical time delays at individual sensors of arrays stations and synthetic waveforms or natural waveforms from similar tectonic environments.

Taking into account the results of our preliminary empirical study one can conclude that cross correlation can significantly reduce the detection threshold for aftershock sequences of nuclear tests worldwide using the IMS seismic network. An additional improvement can be achieved when continuous waveform data from regional stations are available. At regional distances, the duration of correlated signals may be between 10 s (Pn) and 60 s (Lg) which allows gathering substantial integral energy as expressed by higher correlation coefficients. In the best case, the detection threshold might be reduced by an order of magnitude relative to the standard threshold guaranteed by the IMS network.

There are approximately 80,000 events with nonzero depths. However, we would not propose to introduce master events with depth. There are several reasons. First, cross correlation is a tool to search for smaller events, whose depths cannot be determined reliably and therefore fixed to zero in line with the CTBTO's focus on potential explosion. Secondly, nuclear explosions can only be shallow, and zero-depth master events are therefore the best choice for cross correlation. Thirdly, cross correlation may be a useful tool to detect depth dependent phases and thus to constrain the depth. Our limited experience also shows that for such depth-defining phases such as pP and sP, cross correlation could also improve detection and characterization because of their similarity to the primary P-wave. However, this technique works better for events in the lower crust and below when the surface reflected phases are sufficiently separated in time from the primary phase. For shallow events, cross correlation cannot distinguish between P-wave coda and surface reflections.

Cross correlation requires significant resources for real-time computations. With the tentative algorithms and non-optimized software used in this study we have estimated that one standard processor may run forty master events. The regionalized approach allows calculating all master events in parallel and thus the total number of processors is proportional to the number of master events. Under the cross correlation framework, the number of master events has to be estimated iteratively by refining the REB and then XSEL. Previously, we have estimated the number of master events in seismically active areas as ~2,000. This is a very optimistic figure. As a conservative estimate, we would increase the above figure by an order of magnitude and assume the total number of master events as ~20,000, including all isolated events. Then, the number of processors for real-time calculations is ~500. When all algorithms and software are optimized one will likely be able to reduce the number of processors.

**Acknowledgements**
The authors are grateful to their colleagues at the IDC John Coyne, Jeffrey Given and Robert Pearce for help and encouragement. We are also thankful to Kirill Sitnikov and Gadi Turyomurugyendo for building several XSEL events.

**Disclaimer**
The views expressed in this paper are those of the authors and do not necessarily reflect the views of the CTBTO Preparatory Commission.